\documentclass[11pt,a4paper]{article}
\usepackage{natbib}
\usepackage{amsmath,graphicx,color,bm}
\usepackage{amssymb}
\usepackage{amsthm}
\usepackage{multirow}
\usepackage{mathrsfs}
\usepackage{subfigure}
\usepackage{float}
\usepackage{color}
\usepackage{setspace}
\usepackage{hyperref}
\usepackage{appendix}
\usepackage{authblk}
\usepackage{ulem}
\usepackage{algorithm,algorithmic}
\usepackage[top=1in, bottom=1in, left=1.25in, right=1.25in]{geometry}
\usepackage{indentfirst}
\usepackage{fancyvrb}
\usepackage{verbatim}
\newtheorem{theorem}{Theorem}

\newtheorem{corollary}{Corollary}
\newtheorem{remark}{Remark}
\DefineVerbatimEnvironment{verbatim}{Verbatim}{xleftmargin=.0in}

\newcommand{\bx}{\boldsymbol{x}}
\newcommand{\bz}{\boldsymbol{z}}

\begin{document}

\title{Probability-Weighted Clustered Coefficient Regression Models in Complex Survey Sampling}

\date{}
\author[1]{Mingjun Gang}
\author[2]{Xin Wang\thanks{Correspond to: xwang14@sdsu.edu}}
\author[3]{Zhonglei Wang}
\author[3]{Wei Zhong}
\affil[1]{Department of Statistics, National University of Singapore}
\affil[2]{Department of Mathematics and Statistics, San Diego State University}
\affil[3]{MOE Key Laboratory of Econometrics, Wang Yanan Institute for Studies in Economics and School of Economics, Xiamen University}

\maketitle

\begin{abstract}
Regression analysis is commonly conducted in survey sampling. However, existing methods fail when regression models vary across different clusters of domains. In this paper, we propose a unified framework to study the cluster-wise covariate effect under complex survey sampling based on pairwise penalties, and the associated objective function is solved by the alternating direction method of multipliers. Theoretical properties of the proposed method are investigated under regularity conditions. Numerical experiments demonstrate that the proposed method outperforms its alternatives in terms of identifying the cluster structure and estimation efficiency for both linear regression and logistic regression models. American Community Survey is used as an example to illustrate the advantages of the proposed approach.

\end{abstract}

\section{Introduction}
Regression models are commonly used in survey sampling to analyze the relationship among different variables\citep{fuller2011sampling,lohr1994comparison,rao2015small}. Traditional regression models assume common regression coefficients across all domains, but this assumption fails and leads to biased estimators if regression coefficients vary across different clusters of domains. In this paper, we propose a probability-weighted clustered coefficients (PCC) regression model to solve this problem under complex survey sampling.

A concave fusion approach was proposed to identify a cluster structure and estimate model parameters with common regression coefficients but cluster-specific intercepts \cite{ma2017concave}. Extension to models with clustered regression coefficient heterogeneity was investigated by \cite{ma2020exploration}.  Estimation accuracy of \cite{ma2017concave} can be further improved by incorporating spatial information for areal data with repeated measures \cite{wang2019spatial}.  Various models with cluster-specific regression coefficients were explored for different types of data, such as binary data \citep{zhu2021longitudinal}, count data \citep{chen2019subgroup}, survival data \citep {hu2021subgroup}, functional data \citep{wang2023clustering,zhang2022subgroup}, Poisson process \citep{wang2023clustered}; also see \citep{liu2022fusion,zhu2018cluster}. These works have shown the advantages of models with cluster-specific regression coefficients when estimating population parameters. However, none considers identifying a cluster structure of regression coefficients under complex survey sampling. Under complex survey sampling, an empirical-likelihood based method was proposed for variable selection under high dimensional setups \cite{zhao2022sample}, but it cannot be used to identify cluster structures; also see \cite{dumitrescu2021variable,WANG2014409}.


Hierarchical models are widely used in survey sampling and small area estimation \cite{rao2015small}. A multi-level Bayesian model was proposed by \cite{kim2018combining} for small area estimation. A robust data-driven transformation technique was proposed by \cite{Rojas2019}, and a two-level linear regression model with random intercepts was used to obtain a pseudopopulation \cite{molina2010small}. \cite{wang2019small} proposed new model-based estimators using a linear mixed model with cluster-specific regression coefficients, where random effects were considered in the model compared to \cite{ma2020exploration} without random effects. \cite{lahiri2023nested} proposed a model based on linear mixed models with heterogeneity in regression coefficients and variance components. Also see \cite{Ubaidilah2019,datta2000unified,Esteban2022,Ngaruye2017,jiang2006mixed,sun2022bivariate}. Generalized linear models with random effects were also proposed for categorical responses \cite{berg2014small,jiang2001empirical,sun2024multivariate,Zhang2004}. However, existing works did not incorporate sampling weights.  A two-level model with random cluster effects was proposed for population mean estimation \cite{Pfeffermann2007}, and sampling weights are incorporated; also see \cite{marhuenda2017poverty}. Under complex survey sampling, a generalized linear regression model with random effect was considered by \cite{esteban2020small}. The efficiency of existing works can be further improved if we can successfully identify cluster structures, but none pursued in this direction.

Under complex survey sampling, sampling weights are essential to guarantee unbiased inference \citep{pfeffermann1993role},  and statistical efficiency can be improved by incorporating cluster structures. In this work, we consider a PCC  regression model with the smoothly clipped absolute deviation (SCAD) penalty \citep{fan2001variable} to identify the cluster structure by the estimated regression coefficients. We use the alternating direction method of multipliers (ADMM) \citep{boyd2011distributed} to implement the proposed method.  In theory, we show that the oracle estimator, when the cluster structure is available, is a local minimizer of the objective function with probability approaching one under regularity conditions. The asymptotic distribution of our estimator is established accordingly. Theoretical properties are investigated under a general model setup, so they apply to a wide range of situations, including linear and logistic regression models. 

The article is organized as follows. In Section \ref{sec:method}, we propose the PCC regression model incorporating sampling weights and introduce an ADMM-based algorithm to solve the objective function. In Section \ref{sec:thm},  asymptotic properties of the proposed estimators are investigated under regularity conditions. Two simulation studies are conducted to illustrate the advantage of the proposed method for both linear and logistic regression models in Section \ref{sec:sim}. The proposed method is applied to a real survey dataset in Section \ref{sec:app}. Finally, a summary and conclusion are provided in Section \ref{sec:summary}.

\section{Methodology and Algorithm}
\label{sec:method}
\subsection{Basic Setup}
 In this paper, we consider a finite population $\mathcal{F} = \mathcal{F}_1 \cup \cdots \cup \mathcal{F}_m\) of size $N$, where $\{\mathcal{F}_1,\ldots, \mathcal{F}_m\}$ are $m$ mutually exclusive domains,  $\mathcal{F}_i = \{(y_{ih}, \bx_{ih},\bz_{ih}):h=1,\ldots,N_i\}$ for $i=1,\ldots,m$,  $y_{ih}$ is the response of interest for $h$th element in the $i$th domain, 
$\bx_{ih}$ is a \(p\)-dimensional covariate vector associated with the domain-specific part in the regression model, $\bz_{ih}$ is a \(q\)-dimensional covariate vector with respect to the population-level part,  $N_i$ is the size of $\mathcal{F}_i$, and $N = \sum_{i=1}^m N_i$. In this paper, we consider the following generalized linear regression model,
\begin{equation}
    g\left\{E(y_{ih}\mid \bx_{ih},\bz_{ih})\right\} =\boldsymbol{x}_{i h}^{\mathrm{T}} \boldsymbol{\beta}^{0}_{i}+\boldsymbol{z}_{i h}^{\mathrm{T}} \boldsymbol{\eta}^{0} \quad (i =1,...,m; h = 1,..., N_i),\label{eq: generalized regression model} 
\end{equation}
where $g(\cdot)$ is a known link function,  \(\{\boldsymbol{\beta}^{0}_{i}: i = 1,...,m\}\) are  domain-specific regression coefficients, and \(\boldsymbol{\eta}^{0}\) is  common for different domains. For example, $g(x)=x$ corresponds to a  linear regression model, and $g(x)=\log\{x/(1-x)\}$ to a logistic regression model.

 In \eqref{eq: generalized regression model}, \(\{\boldsymbol{\beta}^{0}_{i}: i = 1,...,m\}\) may not be distinct across different domains, and clustering elements with the same domain-specific regression coefficient can effectively improve estimation efficiency. Among \(\{\boldsymbol{\beta}^{0}_{i}: i = 1,...,m\}\),  assume there are $K$ mutually different cluster-specific regression coefficients $\{\boldsymbol{\alpha}_{k}: k=1,\ldots,K\}$, and denote $\mathcal{G}_{k}=\{i:{\boldsymbol{\beta}}_{i}= {\boldsymbol{\alpha}}_{k}, 1 \leq i \leq m\}$ to the domain index set associated with $\boldsymbol{\alpha}_{k}$. In practice, we neither know $\mathcal{G}_k$ nor $K$ in advance, and we are interested in identifying the partition \(\hat{\mathcal{G}}\) and the number of clusters $\hat{K}$, where $\hat{\mathcal{G}}=\{\hat{\mathcal{G}}_{1}, \ldots, \hat{\mathcal{G}}_{\hat{K}}\}$, $\hat{\mathcal{G}}_{k}=\{i: \hat{\boldsymbol{\beta}}_{i}=\hat{\boldsymbol{\alpha}}_{k}, 1 \leq i \leq m\}$, and $\hat{\boldsymbol{\beta}}_{i}$ and $\hat{\boldsymbol{\alpha}}_{k}$ are estimated regression coefficients; see Section~\ref{computation} for details.

If the finite population $\mathcal{F}$ were available, the loss function would be $$ L_N(\bm{\eta},\boldsymbol{\beta})= \sum_{i=1}^{m} W_i L_{i}(\boldsymbol{\eta},\boldsymbol{\beta}),$$ where \(\boldsymbol{\beta} = \left(\boldsymbol{\beta}_{1}^{\mathrm{T}},\ldots,\boldsymbol{\beta}_{m}^{\mathrm{T}}\right)^{\mathrm{T}}\),  $W_i = N_i/N$,  \(L_{i}(\boldsymbol{\eta}, \boldsymbol{\beta}) = N_{i}^{-1} \sum_{h=1}^{N_{i}} L_{ih}(\boldsymbol{\eta}, \boldsymbol{\beta})\), and $L_{ih}(\boldsymbol{\eta},\boldsymbol{\beta})$ is a  loss function corresponding to the $h$th element in the $i$th domain.  For example, $L_{ih}(\bm{\eta},\bm{\beta}) = \frac{1}{2}(y_{ih} - \bm{z}_{ih}^{\mathrm{T}}\bm{\eta} - \bm{x}_{ih}^{\mathrm{T}}\bm{\beta}_i)^2$ corresponds to a linear regression model, and  $L_{ih}(\bm{\eta},\bm{\beta}) =  - y_{ih}\left( \bm{z}_{ih}^{\mathrm{T}}\bm{\eta} + \bm{x}_{ih}^{\mathrm{T}}\bm{\beta}_i\right)  + \log \left\{\exp(\bm{z}_{ih}^{\mathrm{T}}\bm{\eta} + \bm{x}_{ih}^{\mathrm{T}}\bm{\beta}_i) + 1\right\}$ to a logistic regression model.

However, $\mathcal{F}$ is never fully observable in practice due to time and budget constraints. Instead, we can only observe a probability sample. Let  $n_0 = \sum_{i=1}^m n_i$ be the sample size, where $n_i$ is the size with respect to the $i$th domain. Denote \(\delta_{ih}\) as a sampling indicator  with \(\delta_{ih} = 1\) if the \(h\)th element in the \(i\)th domain is observed and \(\delta_{ih} = 0\) otherwise, and let  \(\pi_{ih}\) be the  associated inclusion probability.
Then, a probability-weighted loss function is $L_{\omega}(\boldsymbol{\eta},\boldsymbol{\beta})= \sum_{i=1}^{m} W_i\hat{L}_{i}(\boldsymbol{\eta},\boldsymbol{\beta})$, where $\hat{L}_{i}(\boldsymbol{\eta},\boldsymbol{\beta}) = N_{i}^{-1} \sum_{h=1}^{N_{i}}\delta_{ih}\pi_{ih}^{-1} L_{ih}(\boldsymbol{\eta}, \boldsymbol{\beta})$. It can be shown that $L_{\omega}(\boldsymbol{\eta},\boldsymbol{\beta})$ is design-unbiased for $L_N(\boldsymbol{\eta},\boldsymbol{\beta})$ \citep{horvitz1952generalization}. To estimate the domain-specific regression coefficients $\{\bm{\beta}_i:i=1,\ldots,m\}$ and identify the cluster structure $\mathcal{G}=\{{\mathcal{G}}_{1}, \ldots, {\mathcal{G}}_{{K}}\}$, we consider the following objective function,
\begin{align}Q_{\omega}(\boldsymbol{\eta}, \boldsymbol{\beta}) &=m L_{\omega}( \boldsymbol{\eta},  \boldsymbol{\beta})+\sum_{1\leq i<j \leq m} p_{\gamma}\left(\left\|\boldsymbol{\beta}_{i}-\boldsymbol{\beta}_{j}\right\|, \lambda \right),
\label{mof}
\end{align}
where \(p_{\gamma}\left(\cdot, \lambda \right)\) is a penalty function imposed on all distinct pairs of $\boldsymbol{\beta}_{i}$ and $\boldsymbol{\beta}_{j}$ with $i\neq j$, $\gamma$ is a fixed constant, and $\lambda \geq 0$ is a tuning parameter. In this paper, we use the SCAD penalty with the following form:
\begin{align}
p_{\gamma}(t, \lambda)=\lambda \int_{0}^{|t|} \min \left\{1,(\gamma-x / \lambda)_{+} /(\gamma-1)\right\} \mathrm{d}x, \label{SCAD}
\end{align}
and the tuning parameter is determined by a BIC criterion; see Section~\ref{computation} for details. Following a practical rule \citep{ma2017concave, ma2020exploration,wang2019spatial}, we  fix $\gamma$ to be 3.
There are also other options for the penalty function, such as the minimax concave penalty (MCP)\citep{zhang2010nearly}. \cite{ma2017concave} compared different penalty functions and concluded that SCAD and MCP  perform similarly,  and they are better than an $L_1$ penalty.

\subsection{Algorithm}
\label{computation}

In this section, we use an ADMM-based algorithm to minimize the objective function (\ref{mof}). First, we fix the tuning parameter $\lambda$ and show the algorithm to minimize \eqref{mof}. Details regarding the selection of $\lambda$ are relegated to the end of this section. 

Let  $\boldsymbol{\zeta}_{i j} = \boldsymbol{\beta}_{i}-\boldsymbol{\beta}_{j}$ be the slack parameter associated with $\boldsymbol{\beta}_{i}$ and $\boldsymbol{\beta}_{j}$. Then, the optimization problem is equivalent to minimizing 
\begin{align}
Q_{\omega}(\boldsymbol{\eta},\boldsymbol{\beta}, \boldsymbol{\zeta})= m L_{\omega}(\boldsymbol{\eta},  \boldsymbol{\beta})+\sum_{1 \leq i<j \leq m} p_{\gamma}\left(\left\|\boldsymbol{\zeta}_{i j}\right\|, \lambda\right), \label{obj}\\
\text{subject to} \quad \boldsymbol{\beta}_{i}-\boldsymbol{\beta}_{j}-\boldsymbol{\zeta}_{i j}=\mathbf{0}, \quad 1 \leq i<j \leq m,  \nonumber
\end{align}
where $\boldsymbol{\zeta}=\left(\boldsymbol{\zeta}_{i j}^{\mathrm{T}}, 1 \leq i<j \leq m\right)^{\mathrm{T}}$. The constrained objective function \eqref{obj} can be further formatted as below based on augmented Lagrangian:
\begin{align} Q(\boldsymbol{\eta}, \boldsymbol{\beta}, \boldsymbol{\zeta}, \boldsymbol{v})=Q_{\omega}(\boldsymbol{\eta}, \boldsymbol{\beta}, \boldsymbol{\zeta}) +\sum_{i<j}\left\langle\boldsymbol{v}_{i j}, \boldsymbol{\beta}_{i}-\boldsymbol{\beta}_{j}-\boldsymbol{\zeta}_{i j}\right\rangle+\frac{\vartheta}{2} \sum_{i<j}\left\|\boldsymbol{\beta}_{i}-\boldsymbol{\beta}_{j}-\boldsymbol{\zeta}_{i j}\right\|^{2}, \label{optimization problem}
\end{align}
where $\langle \boldsymbol{a}, \boldsymbol{b}\rangle$ is the inner product of vectors $\boldsymbol{a}$ and $\boldsymbol{b}$, $\boldsymbol{v}=\left(\boldsymbol{v}_{i j}^{\mathrm{T}}, 1 \leq i<j \leq m\right)^{\mathrm{T}}$ are Lagrange multipliers, and $\vartheta > 0$ is a tuning parameter for the penalty. In this paper, we fix \( \vartheta = 1\) following \cite{ma2020exploration}. Then, we use an iterative algorithm to update $\boldsymbol{\beta}, \boldsymbol{\eta}, \boldsymbol{\zeta}$ and $\boldsymbol{v}$. 

Denote $(\boldsymbol{\beta}^{(r)}, \boldsymbol{\eta}^{(r)}, \boldsymbol{\zeta}^{(r)}, \boldsymbol{v}^{(r)})$ as the values at the $r$th iteration, and they can be updated as
 \begin{align}\left(\boldsymbol{\eta}^{(r+1)}, \boldsymbol{\beta}^{(r+1)}\right) &=\underset{\boldsymbol{\eta}, \boldsymbol{\beta}}{\arg \min} \, Q\left(\boldsymbol{\eta}, \boldsymbol{\beta}, \boldsymbol{\zeta}^{(r)}, \boldsymbol{v}^{(r)}\right), \label{eq_step1}\\
\boldsymbol{\zeta}^{(r+1)} &=\underset{\boldsymbol{\zeta}}{\arg \min } \, Q\left(\boldsymbol{\eta}^{(r+1)}, \boldsymbol{\beta}^{(r+1)}, \boldsymbol{\zeta}, \boldsymbol{v}^{(r)}\right), \label{eq_step2}\\
\boldsymbol{v}_{i j}^{(r+1)} &=\boldsymbol{v}_{i j}^{(r)}+\vartheta\left(\boldsymbol{\beta}_{i}^{(r+1)}-\boldsymbol{\beta}_{j}^{(r+1)}-\boldsymbol{\zeta}_{i j}^{(r+1)}\right) \label{eq_v} .\end{align}

For a classical linear regression model, we can derive closed forms for $\boldsymbol{\eta}^{(r+1)}$ and $\boldsymbol{\beta}^{(r+1)}$ by adding sampling weights to the algorithm in \cite{wang2019spatial}; see Appendix \ref{alg_linear} for details. When a logistic regression model is considered, we derive a new algorithm based on the Newton-Raphson method based on \cite{zhu2021longitudinal}; see Appendix \ref{alg_logistic} for details. 

To update $\boldsymbol{\zeta}_{i j}$, it is equivalent to minimizing 
\begin{align}
\frac{\vartheta}{2}\left\|\boldsymbol{\kappa}_{i j}^{(r)}-\boldsymbol{\zeta}_{i j}\right\|^{2}+p_{\gamma}\left(\left\|\boldsymbol{\zeta}_{i j}\right\|, \lambda\right), 
\end{align}
where $\boldsymbol{\kappa}_{i j}^{(r+1)}=(\boldsymbol{\beta}_{i}^{(r+1)}-\boldsymbol{\beta}_{j}^{(r+1)})+\vartheta^{-1} \boldsymbol{v}_{i j}^{(r)}$. For the SCAD penalty, we have  
\begin{equation}
\label{eq:zeta}
    \boldsymbol{\zeta}_{i j}^{(r+1)}= \begin{cases}S(\boldsymbol{\kappa}_{i j}^{(r+1)}, \lambda  / \vartheta) & \text { if }\left\|\boldsymbol{\kappa}_{i j}^{(r+1)}\right\| \leq \lambda +\lambda  / \vartheta, \\ 
    \frac{S\left(\kappa_{i j}^{(r+1)}, \gamma \lambda /(\gamma-1) \vartheta\right)}{1-1 /(\gamma-1) \vartheta} & \text { if } \lambda+\lambda / \vartheta<\left\|\boldsymbol{\kappa}_{i j}^{(r+1)}\right\| \leq \gamma \lambda , \\ \boldsymbol{\kappa}_{i j}^{(r+1)} & \text { if }\left\|\boldsymbol{\kappa}_{i j}^{(r+1)}\right\|>\gamma \lambda ,\end{cases}
\end{equation}
where $\gamma>1+1/ \vartheta$, and $S(\boldsymbol{w}, t)=(1-t /\|\boldsymbol{w}\|)_{+} \boldsymbol{w}$ with $(t)_{+}=t$ if $t>0$ and $0$ otherwise. 

Algorithm~\ref{admm_alg} summarizes the estimation procedure.
\begin{algorithm}[H]
	\caption{The ADMM algorithm}
	\label{admm_alg}
	\begin{algorithmic}[1]
	    \REQUIRE: Initialize $\boldsymbol{\beta}^{(0)}, \boldsymbol{\zeta}^{(0)}$ and $\boldsymbol{v}^{(0)}$ \\
	    \FOR {$r=0,1,2,\dots$}
		\STATE Obtain $\boldsymbol{\beta}^{(r+1)}$ and $\boldsymbol{\eta}^{(r+1)}$ by solving \eqref{eq_step1}.
		\STATE Obtain $\boldsymbol{\zeta}^{(r+1)}$ by \eqref{eq:zeta}.
		\STATE Obtain $\boldsymbol{v}^{(r+1)}$ by \eqref{eq_v}.
		\IF{convergence criterion is met} 
		\STATE{Stop and get the estimates}
		\ELSE
		\STATE { $r=r+1$}
		\ENDIF
		\ENDFOR
	\end{algorithmic}
\end{algorithm}

To find reasonable initial values, we set $ p_{\gamma}(\left\|\boldsymbol{\beta}_{i}-\boldsymbol{\beta}_{j}\right\|;\lambda) =  \lambda \left\|\boldsymbol{\beta}_{i}-\boldsymbol{\beta}_{j}\right\|$ with a small value for  \(\lambda\). Then, let the solution to \eqref{mof} be the initial values for  $\boldsymbol{\beta}^{(0)}$ and $\boldsymbol{\zeta}^{(0)}$.  For a classical linear regression model, the problem becomes a type of ridge regression \cite{ma2020exploration}. For a logistic regression model, estimates can be obtained by \cite{zhu2021longitudinal}. Then, set $\bm{\zeta}_{ij}^{(0)} = \bm{\beta}_i^{(0)} - \bm{\beta}_j^{(0)}$ and $\bm{v}^{(0)} = \bm{0}$.

Denote the solution as 
\begin{align}(\hat{\boldsymbol{\eta}}, \hat{\boldsymbol{\beta}}) &= \underset{\boldsymbol{\eta} \in \mathbb{R}^{q}, \boldsymbol{\beta} \in \mathbb{R}^{mp}}{\arg \min }Q_{\omega}(\boldsymbol{\eta,\beta}),\label{solution}
\end{align}
and $\hat{\boldsymbol{\zeta}}_{i j}=\hat{\boldsymbol{\beta}}_i-\hat{\boldsymbol{\beta}}_j$.
 If $\hat{\boldsymbol{\zeta}}_{i j}=\mathbf{0}$, we conclude that the $i$th and $j$th domains are in the same cluster. Then, the corresponding estimated partition $\hat{\mathcal{G}}$ and the estimated number of clusters $\hat{K}$ are identified automatically. 


\begin{remark}
    The convergence criterion used is the same as that in \cite{ma2017concave} and \cite{wang2019spatial} based on the primal residual $\boldsymbol{r}^{(r+1)}=\boldsymbol{A} \boldsymbol{\beta}^{(r+1)}-\boldsymbol{\zeta}^{(r+1)}.$ 
We stop the algorithm if  $\left\|\boldsymbol{r}^{(r+1)}\right\|<\varepsilon$, where \(\epsilon\) is a small positive number. 
\end{remark}

The tuning parameter $\lambda$ is selected based on modified BIC \citep{wang2007tuning}, which is widely used in clustered coefficient regression models \citep{ma2017concave, wang2019spatial}. Based on the weighted loss function discussed in \cite{lumley2015aic,tang2016fused}, the proposed PCC method selects the tuning parameter $\lambda$ by minimizing
\begin{align}
B I C(\lambda)= \hat{l}(\hat{\bm{\eta}}_\lambda,\hat{\bm{\beta}}_\lambda)+C_{m} \frac{\log n_0}{n_0}
\left\{\hat{K}(\lambda) p\right\}, \label{mod_bic}
\end{align}
where $(\hat{\bm{\eta}_\lambda},\hat{\bm{\beta}}_\lambda)$ are the optimal estimators given $\lambda$, and $C_{m}$ is a positive number depending on the number of domains \(m\). In our paper, we set \(C_{m}=\log (mp+q)\) as in \cite{ma2020exploration}. 
$\hat{l}(\hat{\bm{\eta}},\hat{\bm{\beta}}) = \log(L_{\omega}(\hat{\bm{\eta}}, \hat{\bm{\beta}}))$  for linear regression models and $\hat{l}(\hat{\bm{\eta}},\hat{\bm{\beta}}) = 2L_{\omega}(\hat{\bm{\eta}},\hat{\bm{\beta}})$ for logistic regression models.

\section{Theoretical Properties}
\label{sec:thm}
 Recall that $\mathcal{G}_k$ is the domain index set for the $k$th cluster with cardinality  $|\mathcal{G}_k|$. 
 Denote $\tilde{n}_k= \sum_{i\in \mathcal{G}_k} n_i$ as the number of observations in $\mathcal{G}_k$, and  $n_0 = \sum_{i=1}^m n_i$ as the sample size.
Let $\widetilde{\boldsymbol{W}}=(w_{ik})$ be an $m \times K$ matrix, where $w_{i k}=1$ if $i \in \mathcal{G}_{k}$ and 0 otherwise. Define an $m p \times K p$ matrix $\boldsymbol{W}=\widetilde{\boldsymbol{W}} \otimes \boldsymbol{I}_{p}$, where  \(\boldsymbol{I}_{p}\) is a \(p \times p\) identity matrix, and \(\otimes\) is the Kronecker product. Let $\bm{\alpha} = (\bm{\alpha}_1^T,\dots, \bm{\alpha}_K^T)^T$ be the cluster-specific regression parameters. Then, we have \(\boldsymbol{\beta}= \boldsymbol{W\alpha}\) if the true cluster information is available.  

For simplicity, denote  \(L^{\mathcal{G}}_{ih}\left(\boldsymbol{\eta}, \boldsymbol{\alpha}\right) = L_{ih}\left(\boldsymbol{\eta}, \boldsymbol{W}\boldsymbol{\alpha}\right)\), \(L^{\mathcal{G}}_{i}\left(\boldsymbol{\eta}, \boldsymbol{\alpha}\right) = L_{i}\left(\boldsymbol{\eta}, \boldsymbol{W}\boldsymbol{\alpha}\right)\), \(\hat{L}^{\mathcal{G}}_{i}\left(\boldsymbol{\eta}, \boldsymbol{\alpha}\right) = \hat{L}_{i}\left(\boldsymbol{\eta}, \boldsymbol{W}\boldsymbol{\alpha}\right)\), and \(L^{\mathcal{G}}_{\omega}(\boldsymbol{\eta}, \boldsymbol{\alpha}) = L_{\omega}(\boldsymbol{\eta}, \boldsymbol{W} \boldsymbol{\alpha})\). Let \(\left(\pmb{\eta}^{0},\boldsymbol{\alpha}^{0}\right)\) be the true parameters, and its oracle estimator, the one when the true cluster information is known,  is obtained by
\begin{align}
  \left(\hat{\boldsymbol{\eta}}^{o r}, \hat{\boldsymbol{\alpha}}^{o r}\right) &= \mathop{\arg\min} _{\left(\boldsymbol{\eta}, \boldsymbol{\alpha}\right)  \in  \Theta }L^{\mathcal{G}}_{\omega}(\boldsymbol{\eta}, 
\boldsymbol{\alpha})\notag\\
& = \mathop{\arg\min}_{\left(\boldsymbol{\eta}, \boldsymbol{\alpha}\right)  \in  \Theta }L_{\omega}(\boldsymbol{\eta}, \boldsymbol{W}
\boldsymbol{\alpha})
=\mathop{\arg\min} _{\left(\boldsymbol{\eta}, \boldsymbol{\alpha}\right)  \in  \Theta} \sum_{i=1}^{m}W_iN_{i}^{-1} \sum_{h=1}^{N_{i}}\delta_{ih}\pi_{ih}^{-1}L_{ih}\left(\boldsymbol{\eta}, \boldsymbol{W}\boldsymbol{\alpha}\right),\label{eq: or estm}
\end{align}
and $\hat{\bm{\beta}}^{or} = \bm{W}\hat{\bm{\alpha}}^{or}$, where \(\Theta\) is the parameter space.

More notations are needed to discuss the theoretical properties of the proposed method. Let 
 ${\boldsymbol{S}}_{N}(\boldsymbol{\eta, \alpha}) = \sum_{i=1}^{m} W_i N_{i}^{-1} \sum_{h=1}^{N_{i}}\boldsymbol{S}_{ih}\left(\boldsymbol{\eta}, \boldsymbol{\alpha}\right)$ be the score function, where \(\boldsymbol{S}_{ih}\left(\boldsymbol{\eta}, \boldsymbol{\alpha}\right) =  \partial {L}^{\mathcal{G}}_{ih}\left(\boldsymbol{\eta}, \boldsymbol{\alpha}\right)/ \partial {(\boldsymbol{\eta}^{\mathrm{T}}, \boldsymbol{\alpha}^{\mathrm{T}})}^{\mathrm{T}}\) is with respect to the \(h\)th element in the \(i\)th domain.  Let $\hat{\boldsymbol{S}}_{w}(\boldsymbol{\eta, \alpha}) = \sum_{i=1}^{m}W_i\hat{\boldsymbol{S}}_{i}(\boldsymbol{\eta, \alpha})$ be the sample-level score function, where $\hat{\boldsymbol{S}}_{i}(\boldsymbol{\eta}, \boldsymbol{\alpha})= 
 N_{i}^{-1} \sum_{h=1}^{N_{i}} \delta_{ih}\pi_{ih}^{-1}\boldsymbol{S}_{ih}\left(\boldsymbol{\eta}, \boldsymbol{\alpha}\right)$. It can be shown that $\hat{\boldsymbol{S}}_{w}(\boldsymbol{\eta, \alpha})$ is design-unbiased for ${\boldsymbol{S}}_{N}(\boldsymbol{\eta, \alpha})$. Given the cluster information, the oracle estimator  $\left(\hat{\boldsymbol{\eta}}^{o r}, \hat{\boldsymbol{\alpha}}^{o r}\right)$ solves 
\[\hat{\boldsymbol{S}}_{\omega}(\boldsymbol{\eta, \alpha})={\partial}L^{\mathcal{G}}_{\omega}(\boldsymbol{\eta,\alpha})/ \partial {(\boldsymbol{\eta}^{\mathrm{T}}, \boldsymbol{\alpha}^{\mathrm{T}})}^{\mathrm{T}}
= \sum_{i=1}^{m}W_i\hat{\boldsymbol{S}}_{i}(\boldsymbol{\eta, \alpha}) =\mathbf{0}.\]
Denote ${\boldsymbol{I}}_{i}(\boldsymbol{\eta, \alpha}) = N_{i}^{-1} \sum_{h=1}^{N_{i}}\boldsymbol{I}_{ih}\left(\boldsymbol{\eta}, \boldsymbol{\alpha}\right)$ to  be the information matrix for the $i$th domain, where $\boldsymbol{I}_{ih}\left(\boldsymbol{\eta}, \boldsymbol{\alpha}\right) =-\partial \boldsymbol{S}_{ih}\left(\boldsymbol{\eta}, \boldsymbol{\alpha}\right)/ \partial (\boldsymbol{\eta}^{\mathrm{T}}, \boldsymbol{\alpha}^{\mathrm{T}})$. It can be shown that \(\hat{\boldsymbol{I}}_{i}(\boldsymbol{\eta, \alpha}) = N_{i}^{-1} \sum_{h=1}^{N_{i}}\delta_{ih}\pi_{ih}^{-1}\boldsymbol{I}_{ih}\left(\boldsymbol{\eta}, \boldsymbol{\alpha}\right)\) is design-unbiased for $\boldsymbol{I}_{i}\left(\boldsymbol{\eta}, \boldsymbol{\alpha}\right)$.

The following conditions are required to derive the asymptotic properties of the proposed PCC method.

 \begin{enumerate}
\renewcommand{\labelenumi}{(C\arabic{enumi})}

\item The number of clusters $K$, and the dimensions of covariates $p$ and $q$ are fixed. There exists a positive constant $C_1$ such that  
\begin{eqnarray}
&\max\{N_i:i=1,\ldots,m\}/\min\{N_i:i=1,\ldots,m\}<C_1,\notag \\ 
&\max\{n_i:i=1,\ldots,m\}/\min\{n_i:i=1,\ldots,m\}<C_1,\notag \\ 
&\max\{\tilde{n}_k:k=1,\ldots,K\}/\min\{\tilde{n}_k:k=1,\ldots,K\}<C_1.\notag
\end{eqnarray}
\label{cond:size} 

\item \label{cond:para} The parameter space \(\Theta\) is compact, containing $\left(\boldsymbol{\eta}^{0}, \boldsymbol{\alpha}^{0}\right)$ as an interior point. For $\left(\boldsymbol{\eta}, \boldsymbol{\alpha}\right) \in \Theta,$ $L^{\mathcal{G}}_{ih}\left(\boldsymbol{\eta, \alpha}\right)$ is convex, twice continuously differentiable with respect to $(\boldsymbol{\eta, \alpha})$ and $E\{\lvert L^{\mathcal{G}}_{ih}\left(\boldsymbol{\eta}, \boldsymbol{\alpha}\right)\rvert\}<\infty$ for $i = 1,...,m$ and \(h = 1,...,N_i\). 
Furthermore, $\left(\boldsymbol{\eta}^{0}, \boldsymbol{\alpha}^{0}\right)$ uniquely minimizes $E\{L^{\mathcal{G}}_{ih}\left(\boldsymbol{\eta}, \boldsymbol{\alpha}\right)\}$. 

 \item For $\left(\boldsymbol{\eta}, \boldsymbol{\alpha}\right) \in \Theta$, $\boldsymbol{V}\{L^{\mathcal{G}}_{\omega} \left(\boldsymbol{\eta, \alpha}\right)|\mathcal{F}\} \rightarrow 0$ in probability as   $m \rightarrow \infty $,  where $\boldsymbol{V}\{L^{\mathcal{G}}_{\omega} \left(\boldsymbol{\eta, \alpha}\right)|\mathcal{F}\}$ is the variance of $L_{\omega}^{\mathcal{G}}\left(\boldsymbol{\eta, \alpha}\right)$ conditional on the finite population $\mathcal{F}$. \label{cond:var}

  \item \label{cond:clt}  A central limit theorem holds for the weighted score function $\hat{\boldsymbol{S}}_{\omega} (\boldsymbol{\eta}^{0}, \boldsymbol{\alpha}^{0})$,
  \[
   n_0^{1 / 2} \hat{\boldsymbol{S}}_{\omega} (\boldsymbol{\eta}^{0}, \boldsymbol{\alpha}^{0}) \rightarrow N\left(\mathbf{0}, \boldsymbol{\Sigma}\left(\boldsymbol{\eta}^{0},\boldsymbol{\alpha}^{0}\right)\right)
  \] 
  in distribution as $ n_0 \rightarrow \infty$ with respect to the model (\ref{eq: generalized regression model}) and the sampling mechanism, where $\boldsymbol{\Sigma}\left(\boldsymbol{\eta}^{0},\boldsymbol{\alpha}^{0}\right)$ is positively definitive. 

   \item There exists a compact set $\mathcal{B} \subset \Theta$, such that $\hat{\boldsymbol{I}}_{\omega}(\boldsymbol{\eta, \alpha})$ converges to \(\boldsymbol{\mathcal { I }} (\boldsymbol{\eta}, \boldsymbol{\alpha})\) in probability with respect to the model (\ref{eq: generalized regression model}) and the sampling mechanism uniformly over $\mathcal{B}$ as $ n_{0} \rightarrow \infty$, where   \(\hat{\boldsymbol{I}}_{\omega}(\boldsymbol{\eta, \alpha})=\sum_{i=1}^{m}W_i\hat{\boldsymbol{I}}_{i}(\boldsymbol{\eta, \alpha})\), 
   $ \boldsymbol{\mathcal { I }}(\boldsymbol{\eta}, \boldsymbol{\alpha})$ 
   is the limit of $\sum_{i=1}^{m} W_i E\{{\boldsymbol{I}}_{i}\left(\boldsymbol{\eta}, \boldsymbol{\alpha} \right)\} $ as $m\to\infty$, and $\boldsymbol{\mathcal { I }} (\boldsymbol{\eta}^{0}, \boldsymbol{\alpha}^{0}) $ is invertible. \label{cond:infmat}


\item  There exist two positive constants $C_{2}, C_{3}$ such that $C_{2}<{N}_{i} {n}_{i}^{-1} \pi_{ih}<C_{3}$ for $i = 1,...,m, h=1,..., N_{i}$ almost surely. \label{cond:inclusion}

\item Denote $\rho_{\gamma}(t)=\lambda^{-1} p_{\gamma}(t, \lambda)$ as the scaled penalty function. The function \(\rho_{\gamma}(t)\) is symmetric, non-decreasing, concave on \([0, \infty),\)  and  \(\rho_{\gamma}(0)=0 .\) The value of \(\rho_{\gamma}(t)\) is a constant for \(t \geq a \lambda\) and
 some constant \(a>0\). Its derivative
 \(\rho_{\gamma}^{\prime}(t)\) exists and is continuous except for a finite number of
 values of \(t\), and \(\rho_{\gamma}^{\prime}(0+)=1.\) \label{cond:penalty}

\item  Denote $\bm{y} = (y_{11},\dots, y_{1,N_i},\dots, y_{m,1},\dots, y_{m,N_m})^{\mathrm{T}}$ and $\bm{\mu}(\bm{\eta}^0, \bm{\alpha}^0)$ as the expected value vector of $\bm{y}$. The conditional distribution of $\bm{y}$ given covariates belongs to the canonical exponential family. For any vector $\bm{a}$ and $x>0$, $P(\vert \bm{a}^{\mathrm{T}} \left\{\bm{y} - \bm{\mu}(\bm{\eta}^0,\bm{\alpha}^0)\right\}\vert >\Vert \bm{a}\Vert x)\leq 2\exp(-c_1x^2)$, where $0 <c_1 <\infty$. \label{cond:subgaussian}

\item   There exists a positive constant $c_x <\infty$ such that $\max_{i,h,l} \vert x_{ih,l}\vert \leq c_x$ for $l=1,\dots,p$ and $\max_{i,h,l} \vert z_{ih,l}\vert \leq c_x$  for $l=1,\dots, q$. \label{cond:boundx}
\end{enumerate}

Condition~(C\ref{cond:size}) contains some regularity assumptions for the model setup, and we can show that $W_i\asymp m^{-1}$ and $mn_i\asymp n_0$ for $i=1,\ldots,m$, where $a_n\asymp b_n$ is equivalent to $a_n=O(b_n)$ and $b_n=O(a_n)$. The asymptotic orders of $W_i$ and $n_i$ are required to guarantee the convergence rate in Condition~(C\ref{cond:clt}). By Conditions (C\ref{cond:para}) and (C\ref{cond:var}), the convexity of $ L^{\mathcal{G}}_{ih}\left(\boldsymbol{\eta}, \boldsymbol{\alpha}\right)$ and the pointwise convergence of \(V\{L^{\mathcal{G}}_{\omega} \left(\boldsymbol{\eta, \alpha}\right)|\mathcal{F}\}\) guarantee  \(L^{\mathcal{G}}_{\omega} \left(\boldsymbol{\eta, \alpha}\right) \rightarrow \sum_{i=1}^{m}W_i E\{{L}^{\mathcal{G}}_{i}\left(\boldsymbol{\eta}, \boldsymbol{\alpha}\right)\} \) in probability uniformly   with respect to the model (\ref{eq: generalized regression model}) and the sampling mechanism; see Theorem 10.8 of \citep{rockafellar1970convex} for details. Such uniform convergence guarantees the consistency of the proposed estimators; see Theorem~2.1 of \cite{engle1994handbook} for details.
The twice-continuously differentiable property of \({L}^{\mathcal{G}}_{ih}\left(\boldsymbol{\eta, \alpha}\right)\) guarantees that the weighted score function \(\hat{\boldsymbol{S}}_{i}(\boldsymbol{\eta}, \boldsymbol{\alpha})\) is continuously differentiable, so the mean value theorem applies. 
Condition (C\ref{cond:clt}) assumes a central limit theorem for $\hat{\boldsymbol{S}}_{\omega} (\boldsymbol{\eta}, \boldsymbol{\alpha})$, which is used to derive the asymptotic properties of the oracle estimator and can be verified under general complex sampling designs; see Section \ref{6.4} for details of Poisson sampling and stratified multistage cluster sampling.  In Condition (C\ref{cond:infmat}), the convergence of \(\hat{\boldsymbol{I}}_{\omega}(\boldsymbol{\eta, \alpha})\) is used to derive the limiting distribution of  \(\left(\hat{\boldsymbol{\eta}}^{o r}, \hat{\boldsymbol{\alpha}}^{o r}\right)\) \citep[Theorem 4.1.2]{takeshi1985advanced}. Condition (C\ref{cond:inclusion}) is a regularity condition about inclusion probabilities, and it is used to establish the convergence rate of the estimating equation; see Theorem~1.3.3 of \cite{fuller2011sampling} for details. Conditions (C\ref{cond:penalty})--(C\ref{cond:boundx}) are common assumptions for penalized regression in high-dimensional settings \citep{ma2020exploration,fan2011nonconcave}. 

\begin{remark}
Under the assumption  $m\to\infty$, our theoretical results apply as long as that domain sizes are of the same order, including the case where they are bounded. The analysis can be easily extended to the scenario where $m$ is bounded but the domain sizes diverge at the same rate.  Besides,  we implicitly consider a stratified single-stage sampling design in this paper; see Appendix~\ref{app: multistage cluster sampling}  for a brief discussion under stratified multistage cluster sampling.  
\end{remark}

The following Theorem \ref{lemma 1} and Theorem \ref{th1} show theoretical properties of the oracle estimator in (\ref{eq: or estm}).
  \begin{theorem}
 Under Conditions (C\ref{cond:size})--(C\ref{cond:var}), we have 
$\left(\hat{\boldsymbol{\eta}}^{or}, \hat{\boldsymbol{\alpha}}^{or}\right)
 \rightarrow
 \left(\boldsymbol{\eta}^{0}, \boldsymbol{\alpha}^{0}\right)
$ 
 in probability as $ m \rightarrow \infty$  with respect to the model (\ref{eq: generalized regression model})  and the sampling mechanism. \label{lemma 1}
 \end{theorem}
 
\begin{theorem}
Under Conditions (C\ref{cond:size})-(C\ref{cond:infmat}), we have
 \begin{align}
 n_{0}^{1/2} \left(\left(\hat{\boldsymbol{\eta}}^{or}-\boldsymbol{\eta}^{0}\right)^{\mathrm{T}},\left(\hat{\boldsymbol{\alpha}}^{or}-\boldsymbol{\alpha}^{0}\right)^{\mathrm{T}}\right)^{\mathrm{T}}\rightarrow N \left(\mathbf{0}, \boldsymbol{\Sigma}\right)\label{clt} 
 \end{align}
 in distribution as $m\rightarrow \infty$  with respect to the model (\ref{eq: generalized regression model})  and the sampling mechanism, where $\boldsymbol{\Sigma}=\boldsymbol{\mathcal{I}}\left(\boldsymbol{\eta}^{0}, \boldsymbol{\alpha}^{0}\right)^{-1} \boldsymbol{\Sigma}\left(\boldsymbol{\eta}^{0},  \boldsymbol{\alpha}^{0}\right)\boldsymbol{\mathcal{I}}\left(\boldsymbol{\eta}^{0}, \boldsymbol{\alpha}^{0}\right)^{-1},$ and $\boldsymbol{\mathcal { I }}(\boldsymbol{\eta}, \boldsymbol{\alpha})=\sum_{i=1}^{m} W_i E\{\boldsymbol{I}_{i}\left(\boldsymbol{\eta, \alpha}\right)\}$. 
 \label{th1}
 \end{theorem}
 
The proofs of Theorems \ref{lemma 1}--\ref{th1} are relegated to Appendices~\ref{B.1}--\ref{6.2}, respectively. Theorem \ref{lemma 1} and Theorem \ref{th1} assume the cluster structure is known and consider the theoretical properties of the oracle estimator. Theorem \ref{lemma 1} shows the consistency of \(\boldsymbol{\eta}^{or}\) and \(\boldsymbol{\alpha}^{or}\), and  Theorem \ref{th1} establishes the corresponding limiting distribution. Theorems 1--2 show that the oracle estimator can converge to the true parameters in a rate of $\sqrt{n}_0$.

 
Let $b=\min _{i \in \mathcal{G}_{k}, j \in \mathcal{G}_{k^{\prime}}, k \neq k^{\prime}}\left\|\boldsymbol{\beta}_{i}^{0}-\boldsymbol{\beta}_{j}^{0}\right\|=\min _{k \neq k^{\prime}}\left\|\boldsymbol{\alpha}_{k}^{0}-\boldsymbol{\alpha}_{k^{\prime}}^{0}\right\|$ be the minimal difference between two distinct cluster-specific regression parameters. Theorem \ref{th2} establishes theoretical properties of our proposed estimator in \eqref{solution}.

 \begin{theorem}
  \label{th2}
 Under Conditions~(C\ref{cond:size})--(C\ref{cond:boundx}), suppose \(b>a\lambda\) for some constant $a$, $\lambda \rightarrow 0$ and $n_0^{1/2}\lambda \rightarrow \infty$ as $m\rightarrow \infty$. Then, there exists a
 local minimizer
 \((\hat{\boldsymbol{\eta}}^{\mathrm{T}}, {\hat{\boldsymbol{\beta}}}^{\mathrm{T}})\)
 of the objective function
 \(Q_{\omega}(\boldsymbol{\eta}, \boldsymbol{\beta})\) such that
$ P\{(\hat{\boldsymbol{\eta}}^{\mathrm{T}}, \hat{\boldsymbol{\beta}}^{\mathrm{T}}) =((\hat{\boldsymbol{\eta}}^{o r})^{\mathrm{T}},(\hat{\boldsymbol{\beta}}^{o r})^{\mathrm{T}})\} \rightarrow 1.$

 \end{theorem}
 
 The proof of Theorem \ref{th2} is  in  Appendix~\ref{6.3}. The theoretical properties are investigated under complex survey sampling, which is different from the existing studies.  Theorem~\ref{th2} implies that the true cluster structure can be recovered with probability approaching 1, and that the estimated number of clusters $\hat{K}$ satisfies $P(\hat{K}=K) \rightarrow 1$ under complex survey sampling.

Let $\hat{\bm{\alpha}}$ be the distinct cluster vectors of $\hat{\bm{\beta}}$. According to Theorem \ref{th1} and Theorem \ref{th2}, we have the following result. 

\begin{corollary}
 Suppose the conditions in Theorem \ref{th2} hold, we have
  \begin{align*}
 n_{0}^{1/2} \left(\left(\hat{\boldsymbol{\eta}}-\boldsymbol{\eta}^{0}\right)^{\mathrm{T}},\left(\hat{\boldsymbol{\alpha}}-\boldsymbol{\alpha}^{0}\right)^{\mathrm{T}}\right)^{\mathrm{T}}\rightarrow N \left(\mathbf{0}, \boldsymbol{\Sigma}\right) 
 \end{align*}
 in distribution as $m \rightarrow \infty$.
 \end{corollary}

When $\lambda$ satisfies the conditions in Theorem \ref{th2}, and the cluster structure is recovered with probability approaching 1,  Corollary 1 gives the support for conducting statistical inferences for $\hat{\bm{\alpha}}$.

\section{Simulation Studies}
\label{sec:sim}
In this section, we conduct simulation studies to evaluate the performance of the proposed PCC method. We focus on the performance of identifying the cluster information and only consider cases without the global regression coefficient  $\bm{\eta}$.  We consider two types of models: linear regression models in Section \ref{subsec:sim_linear} and logistic regression models in Section \ref{subsec:sim_logistic}, and compare the proposed  PCC method with the one without using sampling weights, referred to as the ``Clustered Coefficients (CC)'' method. 

To evaluate the cluster performance of the proposed PCC method, we report the estimated cluster number \(\hat K\) and the adjusted Rand index (ARI) \citep{rand1971objective}.  
ARI measures
the degree of agreement between two partitions; the largest value is 1. A larger ARI value indicates a better cluster performance.  We report the average $\hat{K}$ and ARI across  1\,000 simulations, along with standard deviation values in the parentheses. To evaluate the estimation accuracy of $\bm{\beta}$, we report the average root mean square error (RMSE), which is defined as 
\[ \left(\frac{1}{m} \sum_{i=1}^{m}\left\|\hat{\boldsymbol{\beta}}_{i}-\boldsymbol{\beta}_{i}\right\|^{2}\right)^{1/2}.\]


\subsection{Linear regression models}
\label{subsec:sim_linear}

In this section, we consider a linear regression model with regression coefficient heterogeneity. We generate a finite population \(\{(\boldsymbol{x}_{ih}, y_{ih}): i = 1,...,m, h=1,...,H\}\)  with $m=100$ and $H=300$ by
\[y_{i h}=\boldsymbol{x}_{i h}^{\mathrm{T}} \boldsymbol{\beta}_{i}+\epsilon_{i h},\]
where \(\boldsymbol{x}_{i h} = \left (1, x_{ih}\right )^{\mathrm{T}}\), \(x_{ih}\sim N(0,1)\) independently, and conditionally on $x_{ih}$, \(\epsilon_{ih}\) is independently generated from \(N(0,\sigma_{ih}^2)\) with \(\sigma_{ih} = \exp(0.5\vert {\boldsymbol{x}^{\mathrm{T}} _{i h}}\boldsymbol{\beta}_{i}\vert)\). Assume that there are three true clusters, denoted as $\mathcal{G}_1$, $\mathcal{G}_2$ and $\mathcal{G}_3$.
The cluster parameters are  \(\boldsymbol{\beta}_{i}= (-1, -1)^{\mathrm{T}}\) for \(i \in \mathcal{G}_{1} ,\boldsymbol{\beta}_{i}= (0.5, 0.5)^{\mathrm{T}}\)
for \(i \in \mathcal{G}_{2}\) and \(\boldsymbol{\beta}_{i}= (2, 2)^{\mathrm{T}}\) for
\(i \in \mathcal{G}_{3}\). Each domain has an equal probability of being assigned to one of these three clusters.


 We conduct Poisson sampling to generate samples from the finite population with expected sample size \(n_i = 10\), \(n_i = 30\) and \(n_i = 50\).  For  \(i \in \mathcal{G}_{1}\) and \(i \in \mathcal{G}_{3}\), we conduct  Poisson sampling with unequal inclusion probabilities, setting $\pi_{ih}  \propto  \exp(0.3x_{ih})$ for \(i \in \mathcal{G}_{1},\) and $\pi_{ih}  \propto  \exp(0.7x_{ih})$ for \(i \in \mathcal{G}_{3}\). For  \(i \in \mathcal{G}_{2}\), We conduct Poisson sampling with equal inclusion probabilities.

\begin{figure}[t]
    \centering
    \includegraphics[scale=0.6]{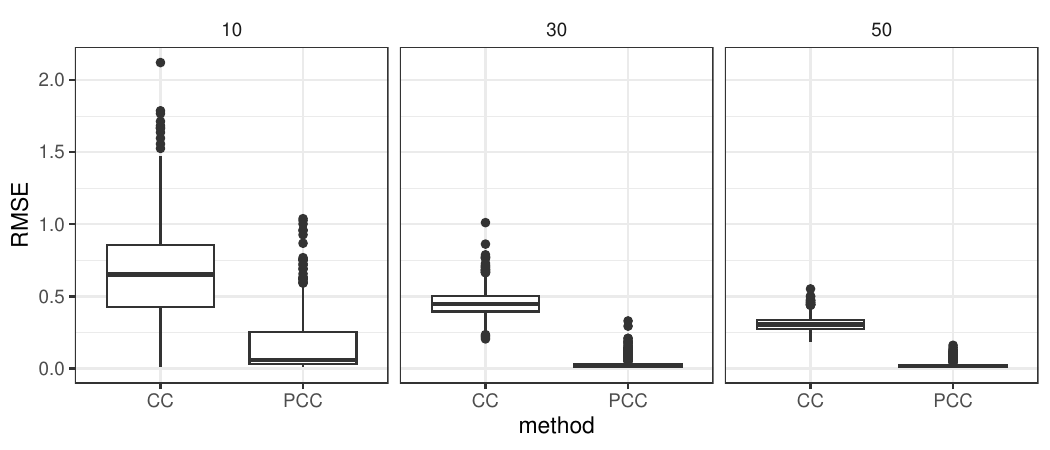}
    \caption{RMSE of $\hat{\bm{\beta}}_{i}$ for the linear regression model based on 1\,000 simulations with expected sample size $ n_i = 10$ (left), $n_i = 30$ (middle) and $n_i=50$ (right).}
    \label{fig:rmse_reg}
\end{figure}

Table~\ref{table_reg} and Figure~\ref{fig:rmse_reg} summarize the results of the linear regression model. For \(n_i =10\), the proposed PCC method has a larger ARI than the CC method. A similar observation occurs when \(n_i = 30\), with the proposed PCC method having the average cluster number closer to the true number of clusters. As $n_i$ increases, both methods have larger ARI values, indicating a better identification of cluster structures. From the results in Figure \ref{fig:rmse_reg}, we can also observe that compared with the CC method, the proposed PCC method provides better estimation accuracy, as indicated by smaller RMSE values.

\begin{table*}[t]
\begin{center}
\caption{Summary of $\hat{K}$ and average ARI  for the linear regression model based on 1\,000 simulations, along with standard deviations in the parentheses. CC and PCC represent the methods without and with sampling weights, respectively. }\label{table_reg}
~\\
\begin{tabular}{ccccc}
\hline &  \multicolumn{2}{c}{$\hat{K}$} & \multicolumn{2}{c}{ ARI } \\
&  CC & PCC & CC & PCC \\

\hline {$n_{i}=10$} & 5.04(1.111) & 3.40(0.582) & 0.94(0.043) & 0.99(0.02) \\ 
{$n_{i}=30$} & 5.00(0.816) & 3.07(0.282) & 0.94(0.048) & 1(0.015) \\ 
$n_i=50$ & 5.13(0.695) & 3.06(0.248) & 0.95(0.045) & 1(0.01) \\ 

\hline
\end{tabular}
\end{center}
\end{table*}

\subsection{Logistic regression models}
\label{subsec:sim_logistic}

In this section, we consider a logistic regression model with regression coefficient heterogeneity. We generate a finite population \(\{(\boldsymbol{x}_{ih}, y_{ih}): i = 1,...,m, h=1,...,H\}\)  with $m=100$ and $H=300$ by
\begin{equation*}
    y_{ih} {\sim} \text{Bernoulli}(p_{ih})
\end{equation*}
independently with $\text{logit}(p_{ih}) = \boldsymbol{x}_{i h}^{\mathrm{T}}\boldsymbol{\beta}_{i}$
where \(\boldsymbol{x}_{i h} = \left (1, x_{ih}\right )^{\mathrm{T}}\), \(x_{ih}\) is simulated from Uniform(-2,2), and $\text{logit}(x) = \log x - \log (1-x)$ for $x\in (0,1)$. As in the preceding section, we assume there are 3 true clusters, denoted as $\mathcal{G}_1$, $\mathcal{G}_2$ and $\mathcal{G}_3$. The cluster parameters are  \(\boldsymbol{\beta}_{i}= (-1, 0.5)^{\mathrm{T}}\) for \(i \in \mathcal{G}_{1} ,\boldsymbol{\beta}_{i}= (0.5, 1.5)^{\mathrm{T}}\)
for \(i \in \mathcal{G}_{2}\) and \(\boldsymbol{\beta}_{i}= (2, -0.5)^{\mathrm{T}}\) for
\(i \in \mathcal{G}_{3}\). We still conduct Poisson sampling to generate samples from the finite population with expected sample size \(n_i = 10\), \(n_i = 30\) and $n_i = 50$.  For  \(i \in \mathcal{G}_{1}\) and \(i \in \mathcal{G}_{3}\), we conduct  Poisson sampling with unequal inclusion probabilities and set $\pi_{ih}  \propto  \exp(0.3x_{ih} + 0.5y_{ih})$ for \(i \in \mathcal{G}_{1},\) and $\pi_{ih}  \propto  \exp(0.7x_{ih} + 0.5y_{ih})$ for \(i \in \mathcal{G}_{3}\). For  \(i \in \mathcal{G}_{2}\), we conduct Poisson sampling with equal inclusion probabilities. 

\begin{figure}[t]
    \centering
    \includegraphics[scale=0.6]{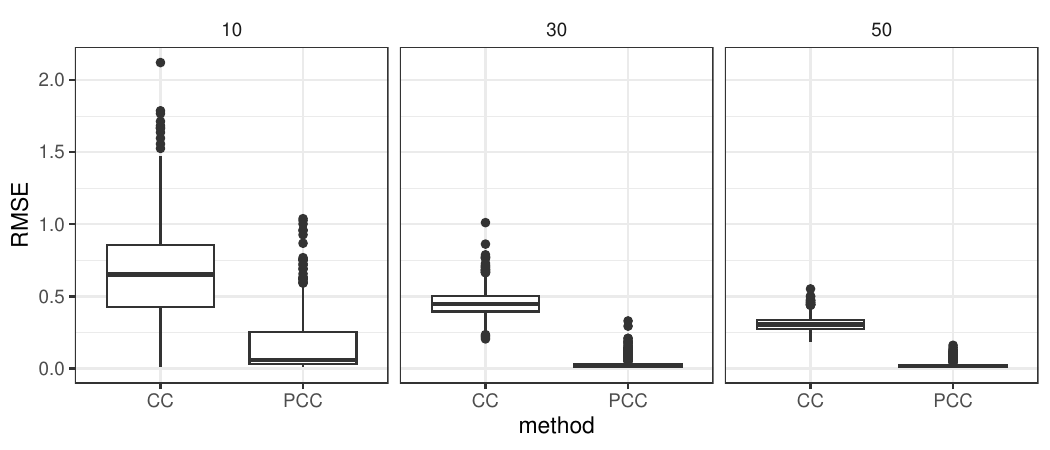}
    \caption{RMSE of $\hat{\bm{\beta}}_{i}$ for the logistic regression model based on 1\,000 simulations with expected sample size $ n_i = 10$ (left), $n_i = 30$ (middle) and $n_i=50$ (right).}
    \label{fig:rmse_logistic}
\end{figure}

Table \ref{table_logistic} shows the results for $\hat{K}$ and ARI, and Figure \ref{fig:rmse_logistic} shows the results for RMSE. It can be seen that when the $n_i=10$, the local sample size is not large, both methods cannot identify the cluster structure well, but the performance of the  proposed PCC method is better. When local sample sizes increase to $n_i=30$, both methods have better results, and the proposed PCC method outperforms the CC method in terms of larger ARI and smaller RMSE values. When $n_i=50$, both methods have similar ARI values, but the estimated number of clusters in PCC is closer to the truth number of clusters, and PCC also has smaller RMSE values. 

\begin{table}[ht]
\begin{center}
\caption{Summary of $\hat{K}$ and average ARI for the logistic regression model based on 1\,000 simulations, along with standard deviations in the parentheses. CC and PCC represent the methods without and with sampling weights, respectively. }\label{table_logistic}
~\\
\begin{tabular}{ccccc}
\hline &  \multicolumn{2}{c}{$\hat{K}$} & \multicolumn{2}{c}{ ARI } \\
&  CC & PCC & CC & PCC \\
\hline {$n_{i}=10$} &  4.46(1.013) & 3.76(0.899) & 0.26(0.135) & 0.36(0.093) \\ 
{$n_{i}=30$} &2.85(1.011) & 3.92(0.939) & 0.63(0.137) & 0.84(0.108) \\ 
$n_i = 50$ & 3.65(0.806) & 3.20(0.46) & 0.95(0.072) & 0.95(0.045) \\ 
\hline
\end{tabular}
\end{center}
\end{table}

 We also consider three additional examples where the datasets are not simulated from the true models in Appendix \ref{sec:more_examples}. In example 1, we simulate data from linear mixed models. We find that when the variance of random effects is large, our PCC can capture the random effects using clustered fixed effects, while CC cannot. In example 2 and example 3, we simulate data from linear regression models with an additional logarithm scale effect. The results show that a large sample size is needed to recover the cluster structure accurately when this additional effect is large. Our proposed PCC still performs better than CC without sampling weights.

\section{Application}
\label{sec:app}
In this section, we apply the proposed PCC method to the American Community Survey (ACS), which is a demographics survey program conducted by the U.S. Census Bureau. This program regularly collects information previously contained only in the long form of the decennial census, such as ancestry, citizenship, educational attainment, income, language proficiency, migration, disability, employment, and housing characteristics. 

The Public Use Microdata Sample (PUMS) files are publicly available data at both individual and household levels.
The most detailed unit of geography contained in the PUMS files is the 
Public Use Microdata Area (PUMA). PUMA contains special non-overlapping areas that partition each state into contiguous geographic units containing roughly 100\,000 people at the time of their creation. 
In this section, we use the household level PUMS data and model the relationship between the logarithm of the monthly gross rent and the logarithm of the household income across 43 PUMAs in Minnesota based on ACS 2021 data. In this dataset, the number of observations in each PUMA ranges from 26 to 163. 


We consider the following model,
$$
y_{i h}=\beta_{0, i}+\beta_{1, i} x_{i h}+\epsilon_{i h},
$$
where \(y_{ih}\) is the logarithm of the monthly gross rent for the \(h\)th household in the \(i\)th PUMA, and \(x_{ih}\) is the logarithm of household income for the \(h\)th household in the \(i\)th PUMA.  We assume there exists coefficient heterogeneity. 


 Tables~\ref{mh_ap}--\ref{sass_ap} provide the estimated regression coefficients in different clusters with standard error in the parenthesis for the CC and PCC methods, respectively. 
 By the CC method, all PUMAs are divided into three clusters.
 All clusters have positive estimates of $\beta_1$. This result implies that the monthly gross is expected to increase as the household income increases in all clusters.  Among these three clusters, the second cluster has the largest $\hat{\beta}_1$. By using the proposed PCC method, four clusters are identified with different estimated regression coefficients.  The estimated coefficients also show a positive relationship between household income and gross rent, which is the strongest in the third cluster, followed by the second cluster, and then the fourth cluster. Table \ref{number} presents a contingency table that displays the number of PUMAs in each cluster. The PUMAs in CC(1) are mainly divided into two clusters by the proposed PCC method. From the estimates in Table \ref{sass_ap}, it can be seen that the estimated regression coefficients of PCC(1) and PCC(2) are quite different, which indicates that it is more reasonable to have two clusters when considering sampling weights.  Furthermore, the PUMAs in CC(3) are separated into PCC(3) and PCC(4), which have different estimates for $\hat{\beta}_0$.

\begin{table}[h!] 
\begin{center}
\caption{Summary of estimated regression coefficients by the CC method.}\label{mh_ap}
~\\
\begin{tabular}{cccc}
\hline cluster & 1 & 2  & 3 \\
\hline
\(\hat{\beta}_0\)& -0.387(0.023) & -0.003(0.029) & 0.347(0.023) \\ 
\(\hat{\beta}_1\) & 0.283(0.022) & 0.62(0.03) & 0.439(0.023) \\ 
\hline
\end{tabular}
\end{center}
\end{table}

\begin{table}[h!]
\begin{center}
\caption{Summary of estimated coefficients by the proposed PCC method.} \label{sass_ap}
~\\
\begin{tabular}{ccccc}
\hline cluster & 1 & 2 & 3  & 4 \\
\hline
\(\hat{\beta}_0\) & -0.292(0.032) & -0.456(0.033) & 0.112(0.021) & 0.479(0.035) \\ 
 \(\hat{\beta}_1\) & 0.201(0.03) & 0.395(0.035) & 0.57(0.021) & 0.353(0.036) \\ 
  
\hline
\end{tabular}
\end{center}
\end{table}

\begin{table}[h!]
\begin{center}
\caption{The number of PUMAs in each cluster for the CC method and the proposed PCC method. CC(\(i\)) represents the \(i\)th cluster in the CC method, where $i = 1,2,3$. PCC(\(j\)) represents the \(j\)th cluster in the proposed PCC method, where \(j = 1,2,3,4.\)  } \label{number}
~\\
\begin{tabular}{ccccc}
\hline  cluster & PCC(1) & PCC(2) & PCC(3)  & PCC(4) \\
\hline
 CC(1)&   7 &   5 &   1 &   0 \\ 
CC(2)  &   0 &   2 &   8 &   0 \\ 
CC(3)  &   0 &   0 &   8 &  12 \\ 
  
\hline
\end{tabular}
\end{center}
\end{table}

\section{Summary and Conclusion}
\label{sec:summary}
 In this paper,  we propose a unified framework (PCC) to estimate regression coefficients as well as identify the cluster structure for domains simultaneously under complex survey sampling. Theoretical properties of the proposed PCC method are investigated under regularity conditions without assuming an explicit form for the model. We also developed algorithms for linear regression and logistic regression models. In the simulation study, the proposed PCC method is compared with a CC method under the setup of a linear regression model and a logistic regression model, and the results demonstrate the importance of sampling weights in identifying the cluster structure under complex survey sampling. Future research can explore new estimators based on probability weighted clustered coefficient regression models for different types of survey data, such as continuous data, binary data, and count data.

%




\begin{appendices}

\section{Computation}
\label{comp}

In this section, the detailed algorithms for updating $\bm{\beta}$ and $\bm{\eta}$ are introduced. In Section \ref{alg_linear}, updates of $\bm{\beta}$ and $\bm{\eta}$ for linear regression models are described. In Section \ref{alg_logistic}, updates of $\bm{\beta}$ and $\bm{\eta}$ for logistic regression models are described.

\subsection{Linear regression models}
\label{alg_linear}

In linear regression model, \(\boldsymbol{\eta}\) and \(\boldsymbol{\beta}\) are updated by the minimizing
\[f(\boldsymbol{\beta}, \boldsymbol{\eta})=\left\|\boldsymbol{\Omega}^{1 / 2}(\boldsymbol{y}-\boldsymbol{Z} \boldsymbol{\eta}-\boldsymbol{X} \boldsymbol{\beta})\right\|^{2}+\vartheta\left\|\boldsymbol{A} \boldsymbol{\beta}-\boldsymbol{\delta}^{(r)}+\vartheta^{-1} \boldsymbol{v}^{(r)}\right\|^{2},\]
where 
\begin{align*}
    &\boldsymbol{y}   =(y_{11},\dots, y_{1 n_{1}},\dots, y_{m 1},\dots, y_{m, n_{m}})^{\mathrm{T}}\\
    &\boldsymbol{Z}  =(\boldsymbol{z}_{11},\dots, \boldsymbol{z}_{1 n_{1}},\dots,\boldsymbol{z}_{m 1},\dots, \boldsymbol{z}_{m, n_{m}})^{\mathrm{T}}\\
    &\boldsymbol{X}  =\operatorname{diag}\left(\boldsymbol{X}_{1},\cdots, \boldsymbol{X}_{m}\right)\\
    &\boldsymbol{\Omega} =m\operatorname{diag}(W_1 / N_{1} \pi_{11}^{-1},\dots W_1 / N_{1} \pi_{1n_{1}} ,\dots,W_m / N_{m} \pi_{m1}^{-1},\dots,W_m / N_{m} \pi_{mn_{m}}^{-1})
\end{align*}
where  $\boldsymbol{X}_{i}=\left(\boldsymbol{x}_{i 1},\dots,\boldsymbol{x}_{i, n_{i}}\right)^{\mathrm{T}}$, $\boldsymbol{A}=\boldsymbol{D} \otimes \boldsymbol{I}_{p}$ with an $m(m-1) / 2 \times m$ matrix $\boldsymbol{D}=\left\{\left(\boldsymbol{e}_{i}-\boldsymbol{e}_{j}\right): 1 \leq i<j \leq m\right\}^{\mathrm{T}}$, \(\otimes\) is the Kronecker product, \(\boldsymbol{I}_p\) is a \(p \times p\)  identity matrix, and $\boldsymbol{e}_{i}$ is an \(m \times 1\) vector with \(i\)th element 1 and other elements 0.
Then, the solutions for \(\boldsymbol{\beta}\) and \(\boldsymbol{\eta}\) are 
\begin{align*}
& \boldsymbol{\beta}^{(r+1)}\\
 = &\left(\boldsymbol{X}^{\mathrm{T}} \boldsymbol{Q}_{Z, \Omega} \boldsymbol{X}+\vartheta \boldsymbol{A}^{\mathrm{T}} \boldsymbol{A}\right)^{-1}\left[\boldsymbol{X}^{\mathrm{T}} \boldsymbol{Q}_{Z, \Omega} \boldsymbol{y}+\vartheta \operatorname{vec}\left\{\left(\boldsymbol{H}^{(r)}-\vartheta^{-1} \boldsymbol{\Upsilon}^{(r)} \right) \boldsymbol{D}\right\}\right], 
\\
& \boldsymbol{\eta}^{(r+1)} = \left(\boldsymbol{Z}^{\mathrm{T}} \boldsymbol{\Omega}\boldsymbol{\Pi}^{-1} \boldsymbol{Z}\right)^{-1} \boldsymbol{Z}^{\mathrm{T}} \boldsymbol{\Omega}\boldsymbol{\Pi}^{-1} \left(\boldsymbol{y}-\boldsymbol{X} \boldsymbol{\beta}^{(r+1)}\right), 
\end{align*}
where $\boldsymbol{Q}_{Z, \Omega}=\boldsymbol{\Omega} - \boldsymbol{\Omega}\boldsymbol{Z}\left(\boldsymbol{Z}^{\mathrm{T}} \boldsymbol{\Omega} \boldsymbol{Z}\right)^{-1} \boldsymbol{Z}^{\mathrm{T}} \boldsymbol{\Omega}$, $\bm{H}^{(r)}=\left\{\boldsymbol{\zeta}_{ij}^{(r)}, 1\leq i<j\leq m\right\}$ and $\bm{\Upsilon}^{(r)}=\left\{\boldsymbol{v}_{ij}^{(r)}: 1\leq i<j \leq m \right\}$. $\bm{H}^{(r)}$ and $\bm{\Upsilon}^{(r)}$ are two matrices with dimension $p \times m(m-1)/2$.


\subsection{Logistic regression models}
\label{alg_logistic}


When updating $\bm{\eta}$ and $\bm{\beta}$, consider the following objective function, 
\[
f\left(\bm{\eta},\bm{\beta}\right)=L_{\omega}(\bm{\eta},\bm{\beta})+\frac{1}{2}\nu\Vert\bm{A}\bm{\beta}-\bm{\delta}^{(r)}+\nu^{-1}\bm{v}^{(r)}\Vert^{2},
\]
where 
\[
L_{\omega}(\bm{\eta},\bm{\beta})=-m\sum_{i=1}^{m}\sum_{h=1}^{n_{i}}\frac{W_i}{N_{i}}\frac{1}{\pi_{ih}}\left(y_{ih}(\bm{z}_{ih}^{T}\bm{\eta}+\bm{x}_{ih}^{T}\bm{\beta}_{i})-\log(1+e^{\bm{z}_{ih}^{T}\bm{\eta}+\bm{x}_{ih}^{T}\bm{\beta}_{i}})\right).
\]

Let $\bm{W}=\text{diag}\left(\bm{w}_{1},\dots,\bm{w}_{m}\right),$ where
$\bm{w}_{i}=\frac{mW_i}{N_{i}}\left(\pi_{i1,},\dots,\pi_{i,n_{i}}\right)$. Denote $\mu_{ih}=\frac{\exp\left(\bm{z}_{ih}^{T}\bm{\eta}+\bm{x}_{ih}^{T}\bm{\beta}_{i}\right)}{1+\exp\left(\bm{z}_{ih}^{T}\bm{\eta}+\bm{x}_{ih}^{T}\bm{\beta}_{i}\right)}$, $\bm{\mu}\left(\bm{\eta},\bm{\beta}\right)=\left(\mu_{ih},i=1\dots,m,h=1,\dots n_{i}\right)$, and  \newline
$\bm{V}=\text{\text{diag}}\left(\bm{\mu}\left(\bm{\eta},\bm{\beta}\right)\left(1-\bm{\mu}\left(\bm{\eta},\bm{\beta}\right)\right)\right)$. Based on the Newton-Raphson algorithm, we can get updates $\bm{\eta}^{(r+1)}$ and $\bm{\beta}^{(r+1)}$ at $s$th inner step as below, 
\begin{align*}
& \bm{\beta}^{\left(r+1,s+1\right)}\\
= & \bm{\beta}^{\left(r+1,s\right)}+\\
 & \left(\bm{X}^{\mathrm{T}}\bm{W}\bm{V}\bm{X}+\nu\bm{A}^{\mathrm{T}}\bm{A}\right)^{-1}\\
 & \left(\bm{X}^{\mathrm{T}}\bm{W}\left(\bm{y}-\bm{\mu}\left(\bm{\eta}^{\left(r+1,s\right)},\bm{\beta}^{\left(r+1,s\right)}\right)\right)-\nu\bm{A}^{\mathrm{T}}\left(\bm{A}\bm{\beta}^{(r+1,s)}-\bm{\delta}^{(r)}+\nu^{-1}\bm{v}^{(r)}\right)\right),
\end{align*}
and 
\[
\bm{\eta}^{\left(r+1,s+1\right)}=\bm{\eta}^{\left(r+1,s\right)}+\left(\bm{Z}^{\mathrm{T}}\bm{W}\bm{V}\bm{Z}\right)^{-1}\bm{Z}^{\mathrm{T}}\bm{W}\left(\bm{y}-\bm{\mu}\left(\bm{\eta}^{\left(r+1,s\right)},\bm{\beta}^{\left(r+1,s\right)}\right)\right).
\]

\section{Proofs of theoretical results}
\label{6.1}
\subsection{Proof of Theorem 1} \label{B.1}
 The proof of Theorem 1 follows Theorem 2.1 in \cite{engle1994handbook}. In this paper, we provide detailed proof of Theorem 1.
 
 Recall that $L_{N}^{\mathcal{G}} \left(\boldsymbol{\eta, \alpha}\right)= \sum_{i=1}^{m}W_iN_{i}^{-1} \sum_{h=1}^{N_{i}}L_{i}^{\mathcal{G}}\left(\boldsymbol{\eta}, \boldsymbol{\alpha}\right)$, where ${L}_{i1}^{\mathcal{G}}\left(\boldsymbol{\eta}, \boldsymbol{\alpha}\right)$ is the loss function for the first element in the \(i\)th unit. By Condition (C\ref{cond:para}) and the weak law of large numbers, we have
 \[ L_{N}^{\mathcal{G}} \left(\boldsymbol{\eta, \alpha}\right) -  \sum_{i=1}^{m}W_iE\{{L}_{i1}^{\mathcal{G}}\left(\boldsymbol{\eta}, \boldsymbol{\alpha}\right)\}\rightarrow 0
 \]
 in probability with respect to the joint distribution of \(\{(\bm{y}_{ih}, \boldsymbol{x}_{ih}, \boldsymbol{z}_{ih}): i = 1,...,m; h = 1,...,N_i\}\). Since $E\{L_{\omega}^{\mathcal{G}} \left(\boldsymbol{\eta, \alpha}\right)| \mathcal{F}\} = L_{N}^{\mathcal{G}} \left(\boldsymbol{\eta, \alpha}\right)$, it follows from Condition (C\ref{cond:var}) that, for $\left(\boldsymbol{\eta}, \boldsymbol{\alpha}\right) \subset \Theta$,
 \[L_{\omega}^{\mathcal{G}} \left(\boldsymbol{\eta, \alpha}\right)-L_{N}^{\mathcal{G}} \left(\boldsymbol{\eta, \alpha}\right) \rightarrow 0\]
 in probability with respect to the sampling mechanism as $m \rightarrow \infty$. Applying Theorem 2 of \cite{jennrich1969asymptotic}, we can show that
 \begin{align}L_{\omega}^{\mathcal{G}} \left(\boldsymbol{\eta, \alpha}\right) -\sum_{i=1}^{m}W_iE\{{L}_{i1}^{\mathcal{G}}\left(\boldsymbol{\eta}, \boldsymbol{\alpha}\right)\} \rightarrow  0 \label{uniform}
 \end{align}
 as $ m \rightarrow \infty$ in probability with respect to the super-population model and the sampling mechanism uniformly over $\Theta$, where $\Theta$ is the overall parameter space. 
 
%

Since $\left(\hat{\boldsymbol{\eta}}^{o r}, \hat{\boldsymbol{\alpha}}^{o r}\right) =\underset{\left(\boldsymbol{\eta}, \boldsymbol{\alpha}\right) \in \Theta }{\arg \min} L_{\omega}^{\mathcal{G}}(\boldsymbol{\eta,\alpha}),$ we have
 \[L_{\omega}^{\mathcal{G}}\left(\hat{\boldsymbol{\eta}}^{o r}, \hat{\boldsymbol{\alpha}}^{o r}\right) \leq L_{\omega}^{\mathcal{G}}(\boldsymbol{\eta}^{0},\boldsymbol{\alpha}^{0}).
 \]
 
%

By the uniform convergence in $\eqref{uniform}$ over $\Theta,$ for any \(\epsilon > 0\), we  have 
 \[|L_{\omega}^{\mathcal{G}}(\boldsymbol{\eta},\boldsymbol{\alpha}) - \sum_{i=1}^{m}W_iE\{{L}_{i1}^{\mathcal{G}}(\boldsymbol{\eta},\boldsymbol{\alpha})\} | < \epsilon/10
 \]
 for sufficiently large $m$ with probability approaching to 1 with respect
  to the super-population model and the sampling mechanism. More specifically, we have the following two results
 \begin{align}
 &\sum_{i=1}^{m}W_iE\{{L}_{i1}^{\mathcal{G}}(\boldsymbol{\eta},\boldsymbol{\alpha})\} < L_{\omega}^{\mathcal{G}}(\boldsymbol{\eta},\boldsymbol{\alpha})+ \epsilon/10, \label{eq1}\\
 &L_{\omega}^{\mathcal{G}}(\boldsymbol{\eta},\boldsymbol{\alpha})<  \sum_{i=1}^{m}W_iE\{{L}_{i1}^{\mathcal{G}}(\boldsymbol{\eta},\boldsymbol{\alpha})\}  + \epsilon/10. \label{eq2}
 \end{align}
 
 
 For any open set $\mathscr{O}$ containing $\left(\boldsymbol{\eta}^{0}, \boldsymbol{\alpha}^{0}\right)$ as an interior point, $\mathscr{O}^{C}\cap \Theta$ is compact. Since $ E\{{L}_{i1}^{\mathcal{G}}\left(\boldsymbol{\eta, \alpha}\right)\}$ is continuous, by extreme value theorem,  we have $\min _{\left(\boldsymbol{\eta}, \boldsymbol{\alpha}\right)  \in \left(\mathscr{O}^{C} \cap \Theta \right)}L_{\omega}^{\mathcal{G}}(\boldsymbol{\eta}, \boldsymbol{\alpha})$  exists.

By Condition (C\ref{cond:para}), $ E\{{L}_{i1}^{\mathcal{G}}\left(\boldsymbol{\eta, \alpha}\right)\}$ is uniquely minimized at the true parameter $\left(\boldsymbol{\eta}^{0}, \boldsymbol{\alpha}^{0}\right).$ We have $$\min _{(\boldsymbol{\eta}, \boldsymbol{\alpha})  \in (\mathscr{O}^{C} \cap \Theta )} \sum_{i=1}^{m}W_iE\{{L}_{i1}^{\mathcal{G}}(\boldsymbol{\eta}, \boldsymbol{\alpha})\}> \sum_{i=1}^m W_i E\{{L}_{i1}^{\mathcal{G}}(\boldsymbol{\eta}^{0}, \boldsymbol{\alpha}^{0})\}.$$ Denote $\epsilon = \min _{\left(\boldsymbol{\eta}, \boldsymbol{\alpha}\right)  \in \left(\mathscr{O}^{C} \cap \Theta \right)} \sum_{i=1}^{m}W_iE\{{L}_{i1}^{\mathcal{G}}\left(\boldsymbol{\eta}, \boldsymbol{\alpha}\right)\}- \sum_{i=1}^{m}W_i E\{{L}_{i1}^{\mathcal{G}}\left({\boldsymbol{\eta}}^{0}, {\boldsymbol{\alpha}}^{0}\right)\}$. By \eqref{eq1} and \eqref{eq2}, we have  
 \[ \begin{aligned}
\min _{\left(\boldsymbol{\eta}, \boldsymbol{\alpha}\right)  \in \left(\mathscr{O}^{C} \cap \Theta \right)}L_{\omega}^{\mathcal{G}}(\boldsymbol{\eta}, \boldsymbol{\alpha}) &\geq \min _{\left(\boldsymbol{\eta}, \boldsymbol{\alpha}\right)  \in \left(\mathscr{O}^{C} \cap \Theta \right)}\sum_{i=1}^{m}W_i E\{{L}_{i1}^{\mathcal{G}}\left({\boldsymbol{\eta}}, {\boldsymbol{\alpha}}\right)\} - \epsilon/10\\
&=\sum_{i=1}^{m}W_iE\{{L}_{i1}^{\mathcal{G}}\left({\boldsymbol{\eta}}^{0}, {\boldsymbol{\alpha}}^{0}\right)\} + \epsilon - \epsilon/10\\
&> L_{\omega}^{\mathcal{G}}(\boldsymbol{\eta}^{0},\boldsymbol{\alpha}^{0}) + 8\epsilon/10\\
&> L_{\omega}^{\mathcal{G}}(\boldsymbol{\eta}^{0},\boldsymbol{\alpha}^{0}) \\
&\geq L_{\omega}^{\mathcal{G}}\left(\hat{\boldsymbol{\eta}}^{o r}, \hat{\boldsymbol{\alpha}}^{o r}\right).
 \end{aligned}\]

 This shows that $\left(\hat{\boldsymbol{\eta}}^{o r}, \hat{\boldsymbol{\alpha}}^{o r}\right) \notin \left(\mathscr{O}^{C} \cap \Theta \right),$ which also means $\left(\hat{\boldsymbol{\eta}}^{o r}, \hat{\boldsymbol{\alpha}}^{o r}\right) \in \mathscr{O}$ with probability approaching 1. Since $\mathscr{O}$ is arbitrary, we can conclude that $\left(\hat{\boldsymbol{\eta}}^{o r}, \hat{\boldsymbol{\alpha}}^{o r}\right)$ is consistent for $\left(\boldsymbol{\eta}^{0}, \boldsymbol{\alpha}^{0}\right)$ by letting $\mathscr{O}$ be small enough. This completes the proof of Theorem \ref{lemma 1}.

 \subsection{Proof of Theorem 2} \label{6.2}
In Theorem 1, we have shown that
 \[\left(\begin{array}{c}
 \hat{\boldsymbol{\eta}}^{o r}\\
 \hat{\boldsymbol{\alpha}}^{o r}
 \end{array}\right)\rightarrow \left(\begin{array}{c}
 \boldsymbol{\eta}^{0}\\
 \boldsymbol{\alpha}^{0}
 \end{array}\right)
 \]
 in probability as \(m \rightarrow \infty\) with respect to superpopulation model and sampling mechanism. By Condition (C\ref{cond:para}), $\hat{\boldsymbol{S}}_{\omega}(\boldsymbol{\eta, \alpha})$ is continuously differentiable.
 Based on the definition of the oracle estimator, we have
 \(\hat{\boldsymbol{S}}_{\omega}(\hat{\boldsymbol{\eta}}^{or}, \hat{\boldsymbol{\alpha}}^{or}) = \boldsymbol{0}.\)
 
 
 By mean value theorem, there exists $(\boldsymbol{\eta}^{\dagger}, \boldsymbol{\alpha}^{\dagger})$ on the segment joining $(\hat{\boldsymbol{\eta}}^{or}, \hat{\boldsymbol{\alpha}}^{or})$ and $({\boldsymbol{\eta}}^{0}, {\boldsymbol{\alpha}}^{0})$ such that 
 \[\begin{aligned}
 \hat{\boldsymbol{S}}_{\omega}({\boldsymbol{\eta}}^{0}, {\boldsymbol{\alpha}}^{0})  &= -\frac{\partial}{\partial (\boldsymbol{\eta}^{\mathrm{T}}, \boldsymbol{\alpha}^{\mathrm{T}})^{\mathrm{T}}}\hat{\boldsymbol{S}}_{\omega}({\boldsymbol{\eta}}^{\dagger}, {\boldsymbol{\alpha}}^{\dagger})\left\{\left((\hat{{\boldsymbol{\eta}}}^{or})^{\mathrm{T}}, (\hat{{\boldsymbol{\alpha}}}^{or})^{\mathrm{T}} \right)^{\mathrm{T}}- \left(({\boldsymbol{\eta}}^{0})^{\mathrm{T}}, ({\boldsymbol{\alpha}}^{0})^{\mathrm{T}}\right)^{\mathrm{T}}\right\}\\
 &= \hat{\boldsymbol{I}}_{\omega}({\boldsymbol{\eta}}^{\dagger}, {\boldsymbol{\alpha}}^{\dagger})\left\{\left((\hat{{\boldsymbol{\eta}}}^{or})^{\mathrm{T}}, (\hat{{\boldsymbol{\alpha}}}^{or})^{\mathrm{T}}\right)^{\mathrm{T}} - \left(({\boldsymbol{\eta}}^{0})^{\mathrm{T}}, ({\boldsymbol{\alpha}}^{0})^{\mathrm{T}}\right)^{\mathrm{T}}\right\}.
 \end{aligned}
 \]
 
 By Condition (C\ref{cond:infmat}) and Theorem 1, we have
 \begin{align}\boldsymbol{\mathcal{I}}\left(\boldsymbol{\eta}^{0}, \boldsymbol{\alpha}^{0}\right)^{-1}{\hat{\boldsymbol{I}}}_{\omega}({\boldsymbol{\eta}}^{\dagger}, {\boldsymbol{\alpha}}^{\dagger}) \rightarrow \boldsymbol{I} \label{CLT_1}
 \end{align}
 in probability with respect to the super-population model and sampling mechanism. Then,
 \begin{align}
 &{n}_{0}^{1/2}\boldsymbol{\mathcal{I}}\left(\boldsymbol{\eta}^{0}, \boldsymbol{\alpha}^{0}\right)^{-1}{\boldsymbol{I}}_{\omega}({\boldsymbol{\eta}}^{\dagger}, {\boldsymbol{\alpha}}^{\dagger})\left\{(\hat{{\boldsymbol{\eta}}}^{or})^{\mathrm{T}}, (\hat{{\boldsymbol{\alpha}}}^{or})^{\mathrm{T}} - (({\boldsymbol{\eta}}^{0})^{\mathrm{T}}, ({\boldsymbol{\alpha}}^{0})^{\mathrm{T}})^{\mathrm{T}}\right\} \notag \\
 &= {n}_{0}^{1/2}\boldsymbol{\mathcal{I}}\left(\boldsymbol{\eta}^{0}, \boldsymbol{\alpha}^{0}\right)^{-1}\hat{\boldsymbol{S}}_{\omega}({\boldsymbol{\eta}}^{0}, {\boldsymbol{\alpha}}^{0}). \label{CLT_2}
 \end{align}
 
 By Conditions (C\ref{cond:clt})–(C\ref{cond:infmat}), $\eqref{CLT_1}, \eqref{CLT_2}$ and Slutsky’s theorem, we have completed the proof of  Theorem \ref{th1}.
 
 \subsection{Proof of Theorem 3} \label{6.3}
 Recall that $L_{\omega}(\boldsymbol{\eta}, \boldsymbol{\beta})$ is the loss function when the cluster information is unknown, and $L_{\omega}^{\mathcal{G}}(\boldsymbol{\eta}, \boldsymbol{\alpha})$ is the loss function when the cluster information is known. In addition, the penalty terms with unknown cluster information and known cluster information have the following forms,
 \[\begin{aligned}
   & P_{m}(\boldsymbol{\beta})=\lambda \sum_{i<j} \rho\left(\left\|\boldsymbol{\beta}_{i}-\boldsymbol{\beta}_{j}\right\|\right) ;\\
 	 &P_{m}^{\mathcal{G}}(\boldsymbol{\alpha})=\lambda \sum_{k<k^{\prime}}\left|\mathcal{G}_{k} \| \mathcal{G}_{k^{\prime}}\right| \rho\left(\left\|\boldsymbol{\alpha}_{k}-\boldsymbol{\alpha}_{k^{\prime}}\right\|\right).
 \end{aligned}\]
  Then, the corresponding objective functions, $Q_{\omega}(\boldsymbol{\eta}, \boldsymbol{\beta}),$ and $ Q_{\omega}^{\mathcal{G}}( \boldsymbol{\eta},  \boldsymbol{\alpha}),$ have the following forms, respectively, 
  \[
  \begin{aligned}
  Q_{\omega}(\boldsymbol{\eta}, \boldsymbol{\beta})= m L_{\omega}( \boldsymbol{\eta},  \boldsymbol{\beta})+P_{m}( \boldsymbol{\beta}), \\
  Q_{\omega}^{\mathcal{G}}( \boldsymbol{\eta},  \boldsymbol{\alpha})= m L_{\omega}^{\mathcal{G}}( \boldsymbol{\eta},  \boldsymbol{\alpha})+P_{m}^{\mathcal{G}}( \boldsymbol{\alpha}).
  \end{aligned}
  \]
  
  Define
  $\mathcal{M}_{\mathcal{G}}=\left\{\boldsymbol{\beta} \in \mathbb{R}^{m p}: \boldsymbol{\beta}_{i}=\boldsymbol{\beta}_{j}, \text{ for } i, j \in \mathcal{G}_{k}, 1 \leq k \leq K\right\}.$
  From the definition, we know that for $\bm{\beta} \in \mathcal{M}_{\mathcal{G}}$, we can write $\bm{\beta} = \bm{W}\bm{\alpha}$. Let 
  \(T: \mathcal{M}_{\mathcal{G}} \rightarrow \mathbb{R}^{K p}\) be the mapping that
  \(T(\boldsymbol{\beta})\) is the \(K p \times 1\) vector consisting of
  \(K\) vectors with dimension \(p\) and its \(k^{\mathrm{th}}\) vector
  component equals to the common value of \(\boldsymbol{\beta}_{i}\) for
  \(i \in \mathcal{G}_{k}\). Let \(T^{*}: \mathbb{R}^{m p} \rightarrow \mathbb{R}^{K p}\) be
  the mapping such that
  \(T^{*}(\boldsymbol{\beta})=\{\left|\mathcal{G}_{k}\right|^{-1} \sum_{i \in \mathcal{G}_{k}} \boldsymbol{\beta}_{i}^{\mathrm{T}}, k=1, \ldots, K\}^{\mathrm{T}}\).
  Clearly, when
  \(\boldsymbol{\beta} \in \mathcal{M}_{\mathcal{G}}, T(\boldsymbol{\beta})=T^{*}(\boldsymbol{\beta}).\)
  
   For every
   \(\boldsymbol{\beta} \in \mathcal{M}_{\mathcal{G}}\), we have
   \(P_{m}(\boldsymbol{\beta})=P_{m}^{\mathcal{G}}(T(\boldsymbol{\beta}))\). And for every \(\boldsymbol{\alpha} \in \mathbb{R}^{Kp}\), we have
   \(P_{m}\left(T^{-1}(\boldsymbol{\alpha})\right)=P_{m}^{\mathcal{G}}(\boldsymbol{\alpha}) .\) Thus, we have 
   $$Q_{\omega}(\boldsymbol{\eta}, \boldsymbol{\beta})=Q_{\omega}^{\mathcal{G}}(\boldsymbol{\eta}, T(\boldsymbol{\beta})),\quad Q_{\omega}^{\mathcal{G}}(\boldsymbol{\eta}, \boldsymbol{\alpha})=Q_{\omega}\left(\boldsymbol{\eta}, T^{-1}(\boldsymbol{\alpha})\right).
   $$

   By Theorem 2, there exists a positive constant $M$, such that
     \[\left\|\hat{\boldsymbol{\eta}}^{o r}-\boldsymbol{\eta}^{0}\right\| \leq M  n_{0}^{-1/2}, \, \max _{i}\left\|\hat{\boldsymbol{\beta}}_{i}^{o r}-\boldsymbol{\beta}_{i}^{0}\right\| \leq M n_{0}^{-1/2}.\]
 
  Denote the neighborhood of \(\left(\boldsymbol{\eta}^{0}, \boldsymbol{\beta}^{0}\right)\) as 
   \[\Theta=\left\{\boldsymbol{\eta} \in \mathbb{R}^{q}, \boldsymbol{\beta} \in \mathbb{R}^{mp}:\left\|\boldsymbol{\eta}-\boldsymbol{\eta}^{0}\right\| \leq Mn_{0}^{-1/2}, \, \max _{i}\left\|\boldsymbol{\beta}_{i}-\boldsymbol{\beta}_{i}^{0}\right\| \leq Mn_{0}^{-1/2}\right\}.\]
   
 Obviously, \(\left(\hat{\boldsymbol{\eta}}^{o r}, \hat{\boldsymbol{\beta}}^{o r}\right) \in \Theta\). For any \(\boldsymbol{\beta} \in \mathbb{R}^{m p}\), let \(\boldsymbol{\beta}^{*}=T^{-1}\left(T^{*}(\boldsymbol{\beta})\right)\).
   We  will show that
   \(\left(\hat{\boldsymbol{\eta}}^{o r}, \hat{\boldsymbol{\beta}}^{o r}\right)\)
   is a strictly local minimizer of the objective function with probability
   approaching 1 through the following two steps.
   
   (i) For any
   \(\left(\boldsymbol{\eta}^{\mathrm{T}}, \boldsymbol{\beta}^{\mathrm{T}}\right)^{\mathrm{T}} \in \Theta\)
   and
  $\left(\boldsymbol{\eta}^{\mathrm{T}},\left(\boldsymbol{\beta}^{*}\right)^{\mathrm{T}}\right)^{\mathrm{T}} \neq
   \left((\hat{\boldsymbol{\eta}}^{or})^{\mathrm{T}},(\hat{\boldsymbol{\beta}}^{o r})^{\mathrm{T}}\right)^{\mathrm{T}},$
   $$Q_{\omega}\left(\boldsymbol{\eta}, \boldsymbol{\beta}^{*}\right)>Q_{\omega}(\hat{\boldsymbol{\eta}}^{o r}, \hat{\boldsymbol{\beta}}^{o r}).$$

   (ii) There is a neighborhood of
   \(\left((\hat{\boldsymbol{\eta}}^{or})^{\mathrm{T}},(\hat{\boldsymbol{\beta}}^{o r})^{\mathrm{T}}\right)^{\mathrm{T}}\),
   denoted by \(\Theta_{m}\) such that for any
   \(\left(\boldsymbol{\eta}^{\mathrm{T}},\boldsymbol{\beta}^{\mathrm{T}}\right)^{\mathrm{T}} \in \Theta_{m} \cap \Theta\)
   and sufficiently large $m$,
   $$Q_{\omega}(\boldsymbol{\eta}, \boldsymbol{\beta}) \geq Q_{\omega}\left(\boldsymbol{\eta}, \boldsymbol{\beta}^{*}\right)
   $$
   
   Therefore, by the results in (i) and (ii), we have
   \(Q_{\omega}(\boldsymbol{\eta}, \boldsymbol{\beta})>Q_{\omega}(\hat{\boldsymbol{\eta}}^{o r}, \hat{\boldsymbol{\beta}}^{o r})\)
   for any
   \(\left(\boldsymbol{\eta}^{\mathrm{T}}, \boldsymbol{\beta}^{\mathrm{T}}\right)^{\mathrm{T}} \in\)
   \(\Theta_{m} \cap \Theta\) and
   \(\left(\boldsymbol{\eta}^{\mathrm{T}},\boldsymbol{\beta}^{\mathrm{T}}\right)^{\mathrm{T}} \neq\left((\hat{\boldsymbol{\eta}}^{o r})^{\mathrm{T}},(\hat{\boldsymbol{\beta}}^{o r})^{\mathrm{T}}\right)^{\mathrm{T}}\), so that
   \(\left((\hat{\boldsymbol{\eta}}^{o r})^{\mathrm{T}},(\hat{\boldsymbol{\beta}}^{o r})^{\mathrm{T}}\right)^{\mathrm{T}}\)
   is a strict local minimizer of
   \(Q_{\omega}(\boldsymbol{\eta}, \boldsymbol{\beta})\) for sufficiently large $m$.
   
   In the following we prove the result in (i). We first show
   \(P_{m}^{\mathcal{G}}\left(T^{*}(\boldsymbol{\beta})\right)=C_{m}\) for
   any \(\boldsymbol{\beta} \in \Theta\), where \(C_{m}\) is a constant
   which does not depend on \(\boldsymbol{\beta}\). Let
   \(T^{*}(\boldsymbol{\beta})=\boldsymbol{\alpha}=\left(\boldsymbol{\alpha}_{1}^{\mathrm{T}}, \ldots, \boldsymbol{\alpha}_{K}^{\mathrm{T}}\right)^{\mathrm{T}}\).
   It suffices to show that
   \(\left\|\boldsymbol{\alpha}_{k}-\boldsymbol{\alpha}_{k^{\prime}}\right\|>a \lambda\)
   for all $ k \neq k^\prime$. Then, by Condition (C\ref{cond:penalty}), 
   \( \rho_{\gamma}\left(\left\|\boldsymbol{\alpha}_{k}-\boldsymbol{\alpha}_{k^{\prime}}\right\|\right)\)
   is a constant, and as a result
   \(P_{m}^{\mathcal{G}}\left(T^{*}(\boldsymbol{\beta})\right)\) is a
   constant. Since
   \[
   \begin{aligned}
   \left\|\boldsymbol{\alpha}_{k}-\boldsymbol{\alpha}_{k^{\prime}}\right\| &=\left\|\boldsymbol{\alpha}_{k}-\boldsymbol{\alpha}_{k}^{0} +\boldsymbol{\alpha}_{k}^{0}-\boldsymbol{\alpha}_{k'}^{0}+\boldsymbol{\alpha}_{k'}^{0}-\boldsymbol{\alpha}_{k^{\prime}}\right\|\\
   &\geq \left\|\boldsymbol{\alpha}_{k}^{0}-\boldsymbol{\alpha}_{k^{\prime}}^{0}\right\|-\left\|\boldsymbol{\alpha}_{k}-\boldsymbol{\alpha}_{k}^{0}+\boldsymbol{\alpha}_{k'}^{0}-\boldsymbol{\alpha}_{k'}\right\|\\
   &\geq\left\|\boldsymbol{\alpha}_{k}^{0}-\boldsymbol{\alpha}_{k^{\prime}}^{0}\right\|-2 \max _{k}\left\|\boldsymbol{\alpha}_{k}-\boldsymbol{\alpha}_{k}^{0}\right\|, 
   \end{aligned}\]
 and we have
  \[\begin{aligned}
 &  \max_{k}\left\|\boldsymbol{\alpha}_{k}-\boldsymbol{\alpha}_{k}^{0}\right\|^{2} \\
  = &\max _{k}\left\|\left|\mathcal{G}_{k}\right|^{-1} \sum_{i \in \mathcal{G}_{k}} \boldsymbol{\beta}_{i}-\boldsymbol{\alpha}_{k}^{0}\right\|^{2}=\max _{k}\left\|\left|\mathcal{G}_{k}\right|^{-1} \sum_{i \in \mathcal{G}_{k}}\left(\boldsymbol{\beta}_{i}-\boldsymbol{\beta}_{i}^{0}\right)\right\|^{2} \\
  = &\max _{k}\left|\mathcal{G}_{k}\right|^{-2}\left\|\sum_{i \in \mathcal{G}_{k}}\left(\boldsymbol{\beta}_{i}-\boldsymbol{\beta}_{i}^{0}\right)\right\|^{2} \leq \max _{k}\left|\mathcal{G}_{k}\right|^{-1} \sum_{i \in \mathcal{G}_{k}}\left\|\left(\boldsymbol{\beta}_{i}-\boldsymbol{\beta}_{i}^{0}\right)\right\|^{2} \\
   \leq & \max _{i}\left\|\boldsymbol{\beta}_{i}-\boldsymbol{\beta}_{i}^{0}\right\|^{2} \leq M^{2}n_{0}^{-1}.
  \end{aligned}
  \]
  
  Then for all \(k\) and \(k^{\prime},\)
  \[\left\|\boldsymbol{\alpha}_{k}-\boldsymbol{\alpha}_{k^{\prime}}\right\| \geq\left\|\boldsymbol{\alpha}_{k}^{0}-\boldsymbol{\alpha}_{k^{\prime}}^{0}\right\|-2 \max _{k}\left\|\boldsymbol{\alpha}_{k}-\boldsymbol{\alpha}_{k}^{0}\right\| \geq b-2 M n_{0} ^{-1/2} >a \lambda,\]
  where the last inequality follows from the assumption that
  \(b>a \lambda \) and \(\lambda \gg n_{0}^{-1/2}\). Therefore, we have
  \(P_{m}^{\mathcal{G}}\left(T^{*}(\boldsymbol{\beta})\right)=C_{m}\).
  
  Since
  \(\left((\hat{\boldsymbol{\eta}}^{o r})^{\mathrm{T}},\left(\hat{\boldsymbol{\alpha}}^{o r}\right)^{\mathrm{T}}\right)^{\mathrm{T}}\)
  is the unique global minimizer of
  \(L_{\omega}^{\mathcal{G}}(\boldsymbol{\eta}, \boldsymbol{\alpha})\), then
  \(L_{\omega}^{\mathcal{G}}\left(\boldsymbol{\eta}, T^{*}(\boldsymbol{\beta})\right)>L_{\omega}^{\mathcal{G}}\left(\hat{\boldsymbol{\eta}}^{o r}, \hat{\boldsymbol{\alpha}}^{o r}\right)\)
  for all
  \(\left(\boldsymbol{\eta}^{\mathrm{T}},\left(T^{*}(\boldsymbol{\beta})\right)^{\mathrm{T}}\right)^{\mathrm{T}} \neq\left(\left(\hat{\boldsymbol{\eta}}^{o r}\right)^{\mathrm{T}},\left(\hat{\boldsymbol{\alpha}}^{o r}\right)^{\mathrm{T}}\right)^{\mathrm{T}}\)
  and hence
  \(Q_{\omega}^{\mathcal{G}}(\boldsymbol{\eta}, T^{*}(\boldsymbol{\beta}))>Q_{\omega}^{\mathcal{G}}(\hat{\boldsymbol{\eta}}^{o r}, \hat{\boldsymbol{\alpha}}^{o r})\)
  for all
  \(T^{*}(\boldsymbol{\beta}) \neq \hat{\boldsymbol{\alpha}}^{o r}\) due to that $P^\mathcal{G}_m(T^*(\bm{\beta})) = P_m^{\mathcal{G}}(\hat{\bm{\alpha}}^{or})$. Since we have
  $Q_{\omega}(\boldsymbol{\eta}, \boldsymbol{\beta})=Q_{\omega}^{\mathcal{G}}(\boldsymbol{\eta}, T(\boldsymbol{\beta}))$  for $\bm{\beta} \in \mathcal{M}_{\mathcal{G}}$ and $Q_{\omega}^{\mathcal{G}}(\boldsymbol{\eta}, \boldsymbol{\alpha})=Q_{\omega}\left(\boldsymbol{\eta}, T^{-1}(\boldsymbol{\alpha})\right)$,
  we have
  $Q_{\omega}^{\mathcal{G}}(\hat{\boldsymbol{\eta}}^{o r}, \hat{\boldsymbol{\alpha}}^{o r})=Q_{\omega}(\hat{\boldsymbol{\eta}}^{o r}, \hat{\boldsymbol{\beta}}^{o r})$
  and
  \(Q_{\omega}^{\mathcal{G}}\left(\boldsymbol{\eta}, T^{*}(\boldsymbol{\beta})\right)=Q_{\omega}\left(\boldsymbol{\eta}, T^{-1}\left(T^{*}(\boldsymbol{\beta})\right)\right)=Q_{\omega}\left(\boldsymbol{\eta}, \boldsymbol{\beta}^{*}\right).\)
   Therefore,
 $$ Q_{\omega}\left(\boldsymbol{\eta}, \boldsymbol{\beta}^{*}\right)> Q_{\omega}(\hat{\boldsymbol{\eta}}^{o r}, \hat{\boldsymbol{\beta}}^{o r})$$
  for all
  \(\boldsymbol{\beta}^{*} \neq \hat{\boldsymbol{\beta}}^{o r}\), and
  the result in (i) is proved.
  
  Next, we prove the result in (ii). 
  
  For a positive sequence \(t_{m}\), let
  \(\Theta_{m}=\left\{\boldsymbol{\beta}_{i}: \max _{i}\left\|\boldsymbol{\beta}_{i}-\hat{\boldsymbol{\beta}}_{i}^{o r}\right\| \leq t_{m}\right\}\).

  For
  \(\left(\boldsymbol{\eta}^{\mathrm{T}}, \boldsymbol{\beta}^{\mathrm{T}}\right)^{\mathrm{T}} \in \Theta_{m} \cap \Theta\),
  by the mean value theorem, we have
  \[Q_{\omega}(\boldsymbol{\eta}, \boldsymbol{\beta})-Q_{\omega}\left(\boldsymbol{\eta}, \boldsymbol{\beta}^{*}\right)=m\Gamma_{1}+\Gamma_{2},\]
  where
  \[\begin{aligned}
  &\Gamma_{1}=\frac{\partial L_{\omega}(\boldsymbol{\eta}, \boldsymbol{\beta}^{s})}{\partial{\boldsymbol{\beta}}^{\mathrm{T}}}\left(\boldsymbol{\beta}-\boldsymbol{\beta}^{*}\right), \\
  &\Gamma_{2}=\sum_{i=1}^{m} \frac{\partial P_{m}\left(\boldsymbol{\beta}^{s}\right)}{\partial \boldsymbol{\beta}_{i}^{\mathrm{T}}}\left(\boldsymbol{\beta}_{i}-\boldsymbol{\beta}_{i}^{*}\right),
  \end{aligned}
  \]
  where \(\boldsymbol{\beta}^{s}=\alpha \boldsymbol{\beta}+(1-\alpha) \boldsymbol{\beta}^{*}\) for some constant \(\alpha \in(0,1)\). 

We know that,
   \[\max _{i}\left\|\boldsymbol{\beta}_{i}^{*}-\boldsymbol{\beta}_{i}^{0}\right\|^{2}=\max_{k}\left\|\boldsymbol{\alpha}_{k}-\boldsymbol{\alpha}_{k}^{0}\right\|^{2} \leq M^{2} n _{0}^{-1}.\]
Since \(\boldsymbol{\beta}_{i}^{s}\) is between
   \(\boldsymbol{\beta}_{i}\) and \(\boldsymbol{\beta}_{i}^{*}\), then,   
 $$
   \begin{aligned}
   \max _{i}\left\|\boldsymbol{\beta}_{i}^{s}-\boldsymbol{\beta}_{i}^{0}\right\| &\leq \alpha \max _{i}\left\|\boldsymbol{\beta}_{i}-\boldsymbol{\beta}_{i}^{0}\right\|+(1-\alpha) \max _{i}\left\|\boldsymbol{\beta}_{i}^{*}-\boldsymbol{\beta}_{i}^{0}\right\| \\&\leq \alpha M n_{0}^{-1/2}+(1-\alpha) M n_{0}^{-1/2}=M n_{0}^{-1/2}.
   \end{aligned}
   $$

We will first show the bound of $\Gamma_1$. 
The partial derivative with respect to \(\boldsymbol{\beta}\) is denoted as $ \tilde{\boldsymbol{S}}(\boldsymbol{\eta}, \boldsymbol{\beta}) = {\partial L_{\omega}(\boldsymbol{\eta}, \boldsymbol{\beta})}/{\partial{\boldsymbol{\beta}}^{\mathrm{T}}}$. Then, we have 
\begin{align*}
  \frac{\partial L_{\omega}(\boldsymbol{\eta}, \boldsymbol{\beta}^{s})}{\partial{\boldsymbol{\beta}^{\mathrm{T}}}} &= \sum_{i=1}^{m}W_iN_{i}^{-1} \sum_{h=1}^{N_{i}} \delta_{ih}\pi_{ih}^{-1} \frac{\partial{L_{ih}(\boldsymbol{\eta}, \boldsymbol{\beta}^{s})}}{ \partial{\boldsymbol{\beta}}^{\mathrm{T}}}\\
  &= \sum_{i=1}^{m} W_i\tilde{\boldsymbol{S}}_{i}(\boldsymbol{\eta}, \boldsymbol{\beta}^{s}),
 \end{align*}
  where $ \tilde{\boldsymbol{S}}_{i}(\boldsymbol{\eta}, \boldsymbol{\beta}^{s})= N_{i}^{-1} \sum_{h=1}^{N_{i}} \delta_{ih}\pi_{ih}^{-1} {\partial{L_{ih}(\boldsymbol{\eta}, \boldsymbol{\beta}^{s})}}/{ \partial{\boldsymbol{\beta}}^{\mathrm{T}}}$. 
By Condition (C\ref{cond:infmat}) and the mean value theorem, there exists $({\tilde{\boldsymbol{\eta}}}, \tilde{\boldsymbol{\beta}})$ lying between $(\boldsymbol{\eta}, \boldsymbol{\beta}^s)$ and $(\boldsymbol{\eta}^{0}, \boldsymbol{\beta}^{0})$ such that $\left\|{\partial^{2}{L_{ih}(\boldsymbol{\eta}, \boldsymbol{\beta})}}/{\partial(\boldsymbol{\eta}^{\mathrm{T}}, \boldsymbol{\beta}^{\mathrm{T}})^{\mathrm{T}}\partial{(\boldsymbol{\eta}^{\mathrm{T}}, \boldsymbol{\beta}^{\mathrm{T}})}}|_{({\tilde{\boldsymbol{\eta}}}, \tilde{\boldsymbol{\beta}})}\right\| \leq C_4$ for sufficiently large \(m\), where \(C_4\) is a positive constant. Combined with Condition (C\ref{cond:inclusion}), we have 
 \begin{align}
 & \max_i\left\| \tilde{\boldsymbol{S}}_{i}(\boldsymbol{\eta}, \boldsymbol{\beta}^{s})-  \tilde{\boldsymbol{S}}_{i}(\boldsymbol{\eta}^{0}, \boldsymbol{\beta}^{0})
 \right\| \nonumber\\
 & \leq \max_i \frac{1}{N_{i}} \sum_{h=1}^{N_{i}} \delta_{ih}\pi_{ih}^{-1} \{ \left\|\partial{L_{ih}(\boldsymbol{\eta}, \boldsymbol{\beta}^{s})} / \partial{\boldsymbol{\beta}}^{\mathrm{T}}- 
 {\partial{L_{ih}(\boldsymbol{\eta}^{0}, \boldsymbol{\beta}^{0})}}/ (\boldsymbol{\beta}^{0})^{\mathrm{T}}\right\|\} \nonumber\\
 &\leq \max_i\frac{1}{N_{i}} \sum_{h=1}^{N_{i}} \delta_{ih}\pi_{ih}^{-1}  \left\|{\frac{\partial^{2}{L_{ih}(\boldsymbol{\eta}, \boldsymbol{\beta})}}{\partial(\boldsymbol{\eta}^{\mathrm{T}}, \boldsymbol{\beta}^{\mathrm{T}})^{\mathrm{T}}\partial{(\boldsymbol{\eta}^{\mathrm{T}}, \boldsymbol{\beta}^{\mathrm{T}})}}\bigg|_{({\tilde{\boldsymbol{\eta}}}, \tilde{\boldsymbol{\beta}})} \{(\boldsymbol{\eta}^{\mathrm{T}}, (\boldsymbol{\beta}^s)^{\mathrm{T}})- ((\boldsymbol{\eta}^{0})^{\mathrm{T}}, (\boldsymbol{\beta}^{0})^{\mathrm{T}})\}}\right\| \nonumber\\
 &\leq  N_{i}^{-1} \sum_{h=1}^{N_{i}} \delta_{ih}\pi_{ih}^{-1} C_{4} M(q+Kp)  n_{0}^{-1/2} \nonumber\\
 &\leq  n_{0} ^{-1/2} N_{i}^{-1} \sum_{h=1}^{N_{i}} \delta_{ih}C_{4} C_{2}^{-1} N_i n_{i}^{-1} M(q+Kp)  \nonumber\\
 &= O_{p}( n_{0} ^{-1/2}). \label{eq:bounds1}
\end{align}

Moreover, it is known that $
\left\|\tilde{\bm{S}}_{i}\left(\bm{\eta}^{0},\bm{\beta}^{0}\right)\right\|\leq\sum_{l=1}^{p}\left|\tilde{\bm{S}}_{i}^{l}\left(\bm{\eta}^{0},\bm{\beta}^{0}\right)\right|,$ where 
$$\tilde{\bm{S}}_{i}^{l}\left(\bm{\eta}^{0},\bm{\beta}^{0}\right)=\frac{1}{N_{i}}\sum_{h=1}^{N_{i}}\delta_{ih}\pi_{ih}^{-1}x_{ih,l}\left(y_{ih}-\mu_{ih}\right).$$ We know that $\frac{1}{N_{i}^{2}}\sum_{h=1}^{N_{i}}\delta_{ih}\pi_{ih}^{-2}x_{ih,l}^{2}\leq\frac{1}{n_{i}^{2}C_{2}}\sum_{h=1}^{N_{i}}\delta_{ih}x_{ih,l}^{2}\leq\frac{c_{x}^{2}}{n_{i}C_{2}}\leq\frac{c_{x}^{2}}{\min_{i}n_{i}C_{2}}$ by Condition (C\ref{cond:boundx}). Then, by Condition (C\ref{cond:subgaussian}), we have 
\begin{align*}&P\left(\left|\tilde{\bm{S}}_{i}^{l}\left(\bm{\eta}^{0},\bm{\beta}^{0}\right)\right|\geq\sqrt{2c_{1}^{-1}C_{2}^{-1}c_{x}^{2}}\sqrt{\frac{\log m}{\min_{i}n_{i}}}\right)\\
= & P\left(\left|\frac{1}{N_{i}}\sum_{h=1}^{N_{i}}\delta_{ih}\pi_{ih}^{-1}x_{ih,l}\left(y_{i}-\mu_{i}\right)\right|\geq\sqrt{2c_{1}^{-1}C_{2}^{-1}c_{x}^{2}\frac{\log m}{\min_{i}n_{i}}}\right)\\
\leq & P\left(\left|\frac{1}{N_{i}}\sum_{h=1}^{N_{i}}\delta_{ih}\pi_{ih}^{-1}x_{ih,l}\left(y_{ih}-\mu_{ih}\right)\right|\geq\sqrt{\frac{1}{N_{i}^{2}}\sum_{h=1}^{N_{i}}\delta_{ih}\pi_{ih}^{-2}x_{ih,l}^{2}}\sqrt{2c_{1}^{-1}\log m}\right)\\
\leq & 2\exp\left(-c_{1}2c_{1}^{-1}\log m\right)=2\exp\left(-2\log m\right)=\frac{2}{m^{2}}.
\end{align*}

Thus, we have 
\begin{align*}
 & P\left(\max_{i}\left\|\tilde{\bm{S}}_{i}\left(\bm{\eta}^{0},\bm{\beta}^{0}\right)\right\|\geq\sqrt{2c_{1}^{-1}C_{2}^{-1}c_{x}^{2}}\sqrt{\frac{\log m}{\min_{i}n_{i}}}\right)\\
\leq & \sum_{i=1}^{m}\sum_{l=1}^{p}P\left(\left|\tilde{\bm{S}^{l}}_{i}\left(\bm{\eta}^{0},\bm{\beta}^{0}\right)\right|\geq\sqrt{2c_{1}^{-1}C_{2}^{-1}c_{x}^{2}}\sqrt{\frac{\log m}{\min_{i}n_{i}}}\right)\\
\leq & \sum_{i=1}^{m}\sum_{l=1}^{p}P\left(\left|\tilde{\bm{S}}_{i}^{l}\left(\bm{\eta}^{0},\bm{\beta}^{0}\right)\right|\geq\sqrt{2c_{1}^{-1}C_{2}^{-1}c_{x}^{2}}\sqrt{\frac{\log m}{\min_{i}n_{i}}}\right)\\
\leq & mp\times\frac{2}{m^{2}}=\frac{2p}{m}.
\end{align*}
The result implies that $\max_{i}\Vert\tilde{\bm{S}}_{i}\left(\bm{\eta}^{0},\bm{\beta}^{0}\right)\Vert=O_{p}\left(\sqrt{\frac{\log m}{\min_{i}n_{i}}}\right)$. Combined with \eqref{eq:bounds1}, we have
   $$\begin{aligned}
&    \max_{i} \left\| \tilde{\boldsymbol{S}}_{i}(\boldsymbol{\eta}, \boldsymbol{\beta}^{s})\right\| \\
= &  \max_{i} \left\|  \tilde{\boldsymbol{S}}_{i}(\boldsymbol{\eta}, \boldsymbol{\beta}^{s}) - \tilde{\boldsymbol{S}}_{i}(\boldsymbol{\eta}^{0}, \boldsymbol{\beta}^{0}) +  \tilde{\boldsymbol{S}}_{i}(\boldsymbol{\eta}^{0}, \boldsymbol{\beta}^{0}) \right\| \\
\leq &  \max_{i} \left\|  \tilde{\boldsymbol{S}}_{i}(\boldsymbol{\eta}, \boldsymbol{\beta}^{s}) - \tilde{\boldsymbol{S}}_{i}(\boldsymbol{\eta}^{0}, \boldsymbol{\beta}^{0})\right\|  + \max_i \left\| \tilde{\boldsymbol{S}}_{i}(\boldsymbol{\eta}^{0}, \boldsymbol{\beta}^{0}) \right\| \\
= & O_{p}\left(\sqrt{\frac{\log m}{\min_{i}n_{i}}}\right).
\end{aligned}$$

  
  Since when
  \(i, j \in \mathcal{G}_{k}, \boldsymbol{\beta}_{i}^{*}=\boldsymbol{\beta}_{j}^{*}\), $\Gamma_1$ can be written as 
   \[ \begin{aligned}
   \Gamma_{1}&= \sum_{k=1}^{K} \sum_{\left\{i, j \in \mathcal{G}_{k}, i<j\right\}} \frac{ \{ {W_i\tilde{\bm{S}}_i(\boldsymbol{\eta}, \boldsymbol{\beta}^{s})}- {W_j\tilde{\bm{S}}_j(\boldsymbol{\eta}, \boldsymbol{\beta}^{s})}\}\left(\boldsymbol{\beta}_{i}-\boldsymbol{\beta}_{j}\right)}{\left|\mathcal{G}_{k}\right|},\\
   \end{aligned}\]

   Then, we have 
   \[\begin{aligned}
& \left\| \frac{ \{ {W_i\tilde{\bm{S}}_i(\boldsymbol{\eta}, \boldsymbol{\beta}^{s})}- {W_j\tilde{\bm{S}}_j(\boldsymbol{\eta}, \boldsymbol{\beta}^{s})}\}\left(\boldsymbol{\beta}_{i}-\boldsymbol{\beta}_{j}\right)}{\left|\mathcal{G}_{k}\right|}\right\| \\
 \leq &  2 \max_i W_i \left|\mathcal{G}_{k}\right|^{-1} \max _{i}\left\|\tilde{\bm{S}}_i(\boldsymbol{\eta}, \boldsymbol{\beta}^{s})\right\| \left\|\boldsymbol{\beta}_{i}-\boldsymbol{\beta}_{j}\right\|\\
 = &  2 \max_i W_i\left|\mathcal{G}_{k}\right|^{-1}  O_{p}\left(\sqrt{\frac{\log m}{\min_{i}n_{i}}}\right)\left\|\boldsymbol{\beta}_{i}-\boldsymbol{\beta}_{j}\right\|.
   \end{aligned} 
   \]

   Based on Condition (C\ref{cond:size}), we know that all $\tilde{n}_k$'s are in the same order and all $n_i$'s are in the same order. Since $K$ is fixed, then $\vert \mathcal{G}_k\vert$ has the same order as $m$.  These imply that $O_p(\sqrt{\frac{\log m}{\min_i n_i}}\vert \mathcal{G}_k\vert^{-1}) = O_p(\sqrt{\frac{\log m}{m}} n_0^{-1/2}).$ Also, combined with Condition (C\ref{cond:size}), then, we have 
\begin{equation}
\label{eq:boundgam1}
  \Gamma_{1} \geq - \frac{1}{m}\sum_{k=1}^{K} \sum_{\left\{i, j \in \mathcal{G}_{k}, i<j\right\}} O_p\left(\sqrt{\frac{\log m}{m}} n_0^{-1/2}\right)  \left\|\boldsymbol{\beta}_{i}-\boldsymbol{\beta}_{j}\right\| .  
\end{equation}

   Next we consider $\Gamma_{2}.$
 $$
   \begin{aligned}
   \Gamma_{2}=& \lambda \sum_{\{j>i\}} \rho_{\gamma}^{\prime}\left(\left\|\boldsymbol{\beta}_{i}^{s}-\boldsymbol{\beta}_{j}^{s}\right\|\right)\left\|\boldsymbol{\beta}_{i}^{s}-\boldsymbol{\beta}_{j}^{s}\right\|^{-1}\left(\boldsymbol{\beta}_{i}^{s}-\boldsymbol{\beta}_{j}^{s}\right)^{\mathrm{T}}\left(\boldsymbol{\beta}_{i}-\boldsymbol{\beta}_{i}^{*}\right) \\
   &+\lambda \sum_{\{j<i\}} \rho_{\gamma}^{\prime}\left(\left\|\boldsymbol{\beta}_{i}^{s}-\boldsymbol{\beta}_{j}^{s}\right\|\right)\left\|\boldsymbol{\beta}_{i}^{s}-\boldsymbol{\beta}_{j}^{s}\right\|^{-1}\left(\boldsymbol{\beta}_{i}^{s}-\boldsymbol{\beta}_{j}^{s}\right)^{\mathrm{T}}\left(\boldsymbol{\beta}_{i}-\boldsymbol{\beta}_{i}^{*}\right) \\
   =& \lambda \sum_{\{j>i\}} \rho_{\gamma}^{\prime}\left(\left\|\boldsymbol{\beta}_{i}^{s}-\boldsymbol{\beta}_{j}^{s}\right\|\right)\left\|\boldsymbol{\beta}_{i}^{s}-\boldsymbol{\beta}_{j}^{s}\right\|^{-1}\left(\boldsymbol{\beta}_{i}^{s}-\boldsymbol{\beta}_{j}^{s}\right)^{\mathrm{T}}\left(\boldsymbol{\beta}_{i}-\boldsymbol{\beta}_{i}^{*}\right) \\
   &+\lambda \sum_{\{i<j\}} \rho_{\gamma}^{\prime}\left(\left\|\boldsymbol{\beta}_{j}^{s}-\boldsymbol{\beta}_{i}^{s}\right\|\right)\left\|\boldsymbol{\beta}_{j}^{s}-\boldsymbol{\beta}_{i}^{s}\right\|^{-1}\left(\boldsymbol{\beta}_{j}^{s}-\boldsymbol{\beta}_{i}^{s}\right)^{\mathrm{T}}\left(\boldsymbol{\beta}_{j}-\boldsymbol{\beta}_{j}^{*}\right) \\
   =& \lambda \sum_{\{j>i\}} \rho_{\gamma}^{\prime}\left(\left\|\boldsymbol{\beta}_{i}^{s}-\boldsymbol{\beta}_{j}^{s}\right\|\right)\left\|\boldsymbol{\beta}_{i}^{s}-\boldsymbol{\beta}_{j}^{s}\right\|^{-1}\left(\boldsymbol{\beta}_{i}^{s}-\boldsymbol{\beta}_{j}^{s}\right)^{\mathrm{T}}\left\{\left(\boldsymbol{\beta}_{i}-\boldsymbol{\beta}_{i}^{*}\right)-\left(\boldsymbol{\beta}_{j}-\boldsymbol{\beta}_{j}^{*}\right)\right\}.
   \end{aligned}
 $$
 
   When
   \(i, j \in \mathcal{G}_{k}, \boldsymbol{\beta}_{i}^{*}=\boldsymbol{\beta}_{j}^{*}\),
   and
   \(\boldsymbol{\beta}_{i}^{s}-\boldsymbol{\beta}_{j}^{s}=\alpha\left(\boldsymbol{\beta}_{i}-\boldsymbol{\beta}_{j}\right)\). Thus,
 $$
   \begin{aligned}
   \Gamma_{2}&= \lambda \sum_{k=1}^{K} \sum_{\left\{i, j \in \mathcal{G}_{k}, i<j\right\}} \rho_{\gamma}^{\prime}\left(\left\|\boldsymbol{\beta}_{i}^{s}-\boldsymbol{\beta}_{j}^{s}\right\|\right)\left\|\boldsymbol{\beta}_{i}^{s}-\boldsymbol{\beta}_{j}^{s}\right\|^{-1}\left(\boldsymbol{\beta}_{i}^{s}-\boldsymbol{\beta}_{j}^{s}\right)^{\mathrm{T}}\left(\boldsymbol{\beta}_{i}-\boldsymbol{\beta}_{j}\right) \\
   &+\lambda \sum_{k<k^{\prime}} \sum_{\left\{i \in \mathcal{G}_{k}, j^{\prime} \in \mathcal{G}_{k^{\prime}}\right\}} \rho_{\gamma}^{\prime}\left(\left\|\boldsymbol{\beta}_{i}^{s}-\boldsymbol{\beta}_{j}^{s}\right\|\right)\left\|\boldsymbol{\beta}_{i}^{s}-\boldsymbol{\beta}_{j}^{s}\right\|^{-1} \\
   & \cdot \left(\boldsymbol{\beta}_{i}^{s}-\boldsymbol{\beta}_{j}^{s}\right)^{\mathrm{T}}\left\{\left(\boldsymbol{\beta}_{i}-\boldsymbol{\beta}_{i}^{*}\right)-\left(\boldsymbol{\beta}_{j}-\boldsymbol{\beta}_{j}^{*}\right)\right\}.
   \end{aligned}
   $$

   Hence, for
   \(k \neq k^{\prime}, i \in \mathcal{G}_{k}, j^{\prime} \in \mathcal{G}_{k^{\prime}},\)
   \[\left\|\boldsymbol{\beta}_{i}^{s}-\boldsymbol{\beta}_{j}^{s}\right\| \geq \min _{i \in \mathcal{G}_{k}, j^{\prime} \in \mathcal{G}_{k^{\prime}}}\left\|\boldsymbol{\beta}_{i}^{0}-\boldsymbol{\beta}_{j}^{0}\right\|-2 \max _{i}\left\|\boldsymbol{\beta}_{i}^{s}-\boldsymbol{\beta}_{i}^{0}\right\| \geq b-2 M n_{0}^{-1/2}>a \lambda,\]
   
   and thus
   \(\rho_{\gamma}^{\prime}\left(\left\|\boldsymbol{\beta}_{i}^{s}-\boldsymbol{\beta}_{j}^{s}\right\|\right)=0\) by Condition (C\ref{cond:penalty}). Therefore,
 $$
   \begin{aligned}
   \Gamma_{2} &=\lambda \sum_{k=1}^{K} \sum_{\left\{i, j \in \mathcal{G}_{k}, i<j\right\}} \rho_{\gamma}^{\prime}\left(\left\|\boldsymbol{\beta}_{i}^{s}-\boldsymbol{\beta}_{j}^{s}\right\|\right)\left\|\boldsymbol{\beta}_{i}^{s}-\boldsymbol{\beta}_{j}^{s}\right\|^{-1}\left(\boldsymbol{\beta}_{i}^{s}-\boldsymbol{\beta}_{j}^{s}\right)^{\mathrm{T}}\left(\boldsymbol{\beta}_{i}-\boldsymbol{\beta}_{j}\right) \\
   &=\lambda \sum_{k=1}^{K} \sum_{\left\{i, j \in \mathcal{G}_{k}, i<j\right\}} \rho_{\gamma}^{\prime}\left(\left\|\boldsymbol{\beta}_{i}^{s}-\boldsymbol{\beta}_{j}^{s}\right\|\right)\left\|\boldsymbol{\beta}_{i}-\boldsymbol{\beta}_{j}\right\|,
   \end{aligned}
 $$\\
   where the last step follows from
   \(\boldsymbol{\beta}_{i}^{s}-\boldsymbol{\beta}_{j}^{s}=\alpha\left(\boldsymbol{\beta}_{i}-\boldsymbol{\beta}_{j}\right)\).
   
  When $i,j \in \mathcal{G}_k$, we have $\bm{\beta}_i^* = \bm{\beta}_j^*$ and 
   \[\max _{i}\Vert\boldsymbol{\beta}_{i}^{*}-\widehat{\boldsymbol{\beta}}_{i}^{o r}\Vert=\max _{k}\Vert \boldsymbol{\alpha}_{k}-\widehat{\boldsymbol{\alpha}}_{k}^{o r}\Vert \leq \max _{i}\Vert\boldsymbol{\beta}_i-\widehat{\boldsymbol{\beta}}_{i}^{o r}\Vert.\] Then, 
 $$
   \begin{aligned}
   \max _{i}\left\|\boldsymbol{\beta}_{i}^{s}-\boldsymbol{\beta}_{j}^{s}\right\| & \leq 2 \max _{i}\Vert\boldsymbol{\beta}_{i}^{s}-\boldsymbol{\beta}_{i}^{*}\Vert \leq 2 \max _{i}\Vert\boldsymbol{\beta}_{i}-\boldsymbol{\beta}_{i}^{*}\Vert \\
   & \leq 2\left(\max _{i}\Vert\boldsymbol{\beta}_{i}-\widehat{\boldsymbol{\beta}}_{i}^{o r}\Vert+\max _{i}\Vert\boldsymbol{\beta}_{i}^{*}-\widehat{\boldsymbol{\beta}}_{i}^{o r}\Vert\right) \\&\leq 4 \max _{i}\Vert\boldsymbol{\beta}_{i}-\widehat{\boldsymbol{\beta}}_{i}^{o r}\Vert \\&\leq 4 t_{m}.
   \end{aligned}
 $$
   Hence
   \(\rho_{\gamma}^{\prime}\left(\left\|\boldsymbol{\beta}_{i}^{s}-\boldsymbol{\beta}_{j}^{s}\right\|\right) \geq \rho_{\gamma}^{\prime}\left(4 t_{m}\right)\)
   by concavity of \(\rho_{\gamma}(\cdot)\).  As a result,   
   \begin{equation}
   \label{eq:boundgam2}
       \Gamma_{2} \geq \sum_{k=1}^{K} \sum_{\left\{i, j \in \mathcal{G}_{k}, i<j\right\}} \lambda \rho^{\prime}_{\gamma}\left(4 t_{m}\right)\left\|\boldsymbol{\beta}_{i}-\boldsymbol{\beta}_{j}\right\|.
   \end{equation}

Based on \eqref{eq:boundgam1} and \eqref{eq:boundgam2}, we have
\begin{align*}
 & Q_{\omega}(\boldsymbol{\eta}, \boldsymbol{\beta})-Q_{\omega}\left(\boldsymbol{\eta}, \boldsymbol{\beta}^{*}\right) \\
 \geq & \sum_{k=1}^{K} \sum_{\left\{i, j \in \mathcal{G}_{k}, i<j\right\}}\left\{\lambda  \rho_{\gamma}^{\prime}\left(4 t_{m}\right)- O_p\left(\sqrt{\frac{\log m}{m}} n_0^{-1/2}\right)\right\}\left\|\boldsymbol{\beta}_{i}-\boldsymbol{\beta}_{j}\right\|.
\end{align*}

As \(t_{m}=o(1), \rho_{\gamma}^{\prime}\left(4 t_{m}\right) \rightarrow 1\) by Condition (C\ref{cond:penalty}). Since $\lambda \gg  n_0^{-1/2}$, therefore,
\(Q_{\omega}(\boldsymbol{\eta}, \boldsymbol{\beta}) > Q_{\omega}\left(\boldsymbol{\eta}, \boldsymbol{\beta}^{*}\right) .\) The result in (ii) is proved.
   
   By combining (i) and (ii), we will have that
   $Q_{\omega}(\boldsymbol{\eta}, \boldsymbol{\beta})>Q_{\omega}(\hat{\boldsymbol{\eta}}^{o r}, \hat{\boldsymbol{\beta}}^{o r})$
   for any
   \(\left(\boldsymbol{\eta}^{\mathrm{T}}, \boldsymbol{\beta}^{\mathrm{T}}\right)^{\mathrm{T}} \in \Theta_{n} \cap \Theta\)
   and
   \(\left(\boldsymbol{\eta}^{\mathrm{T}}, \boldsymbol{\beta}^{\mathrm{T}}\right)^{\mathrm{T}} \neq\left((\hat{\boldsymbol{\eta}}^{o r})^{\mathrm{T}},(\hat{\boldsymbol{\beta}}^{o r})^{\mathrm{T}}\right)^{\mathrm{T}} .\)
   This shows that
   \(\left((\hat{\boldsymbol{\eta}}^{o r})^{\mathrm{T}},(\hat{\boldsymbol{\beta}}^{o r})^{\mathrm{T}}\right)^{\mathrm{T}}\)
   is a strict local minimizer of the objective function for
   sufficiently large $m$.

\section{Conditions verification}
\label{6.4}
 \subsection{Poisson Sampling} 
 To demonstrate the rationality of the proposed method, we verify the generality conditions in Section~\ref{sec:thm} for the case where Poisson sampling is conducted for each domain; see \cite{kim2023} for a similar treatment.  In this section, we verify conditions specifically for Poisson sampling, including (C\ref{cond:var})--(C\ref{cond:infmat}). We now assume the following specific conditions.
 
 \begin{enumerate}
 \renewcommand{\labelenumi}{(A\arabic{enumi})}
\item The parameter space \(\Theta\) is compact, containing $\left(\boldsymbol{\eta}^{0}, \boldsymbol{\alpha}^{0}\right)$ as an interior point. For $\left(\boldsymbol{\eta}, \boldsymbol{\alpha}\right) \subset \Theta,$ $L^{\mathcal{G}}_{ih}\left(\boldsymbol{\eta, \alpha}\right)$ is convex and twice continuously differentiable with respect to $(\boldsymbol{\eta, \alpha})$ for $i = 1,...,m$ and \(h = 1,...,N_i\), where $m\to\infty$. 
Besides, $E[L^{\mathcal{G}}_{ih}\left(\boldsymbol{\eta, \alpha}\right)]^2<C_0$ for a constant $C_0$. $\left(\boldsymbol{\eta}^{0}, \boldsymbol{\alpha}^{0}\right)$ uniquely minimizes $E\{L^{\mathcal{G}}_{ih}\left(\boldsymbol{\eta}, \boldsymbol{\alpha}\right)\}$. \label{Poi: true parameter}
\item The pair $\left(\pi_{ih},\bx_{ih},\bz_{ih} y_{ih}\right)$ is independent of $\left(\pi_{jk},\bx_{jk},\bz_{jk}, y_{jk}\right)$ for $i \neq j$ or $h \neq k.$ \label{Poi: inc prob}
\item For all $i = 1,...,m,$ there exist two constants $0<C_{1}<C_{2}<\infty$ such that $C_{1}<{N}_{i} \bar{n}_{i}^{-1} \pi_{ih}<C_{2}$ for $h=1, \ldots, N_{i}$ almost surely with respect to the super-population model, where $\bar{n}_{i}=\sum_{h=1}^{N_{i}} \pi_{ih}$ is the expected sample size in unit $i$ satisfying $\bar{n}_{i}=o_p(N_i)$, $\max\{\bar{n}_{i}:i=1,\ldots,m\}/\min\{\bar{n}_{i}:i=1,\ldots,m\}<C_3$, and $C_3$ is a positive constant. In addition,
\begin{align*}
    & \max\{{N}_{i}:i=1,\ldots,m\}/\min\{{N}_{i}:i=1,\ldots,m\}<C_3,\\
    & \max\{\tilde{n}_{k}:k=1,\ldots,K\}/\min\{\tilde{n}_{k}:k=1,\ldots,K\}<C_3,
\end{align*}
where $\tilde{n}_k$ is the sample size of the $k$th cluster. \label{Poi: exp ss} 
 
\item There exists a compact set $\mathcal{B} \subset \Theta$ containing $(\boldsymbol{\eta}^{0}, \boldsymbol{\alpha}^0)$ as an interior point, such that $\sup _{(\boldsymbol{\eta}, \boldsymbol{\alpha}) \in \mathcal{B}} E\|\hat{\boldsymbol{S}}_i\left(\boldsymbol{\eta}, \boldsymbol{\alpha}\right)\|^{4}<\infty$, where $\|\boldsymbol{x}\|$ is the Euclidean norm of a vector $\boldsymbol{x}$.\label{Poi: estimating function}
\item Given any $\boldsymbol{a} \in \mathbb{R}^{q+Kp}$ satisfying $\| \boldsymbol{a}\|= 1, V\{S_{\boldsymbol{a},ih}(\boldsymbol{\eta}^0, \boldsymbol{\alpha}^0)\} > 0$, and there exists a positive definite matrix $\boldsymbol{\Sigma}_{a}$ such that
 $$
 \sum_{i=1}^{m}W_i^2\bar{n}_{i} N_{i}^{-2} \sum_{h=1}^{N_i} \pi_{ih}^{-1}\left(1, S_{\boldsymbol{a}, ih}\left(\boldsymbol{\eta}^0, \boldsymbol{\alpha}^0\right)\right)^{\mathrm{T}}\left(1, S_{\boldsymbol{a}, ih}\left(\boldsymbol{\eta}^0, \boldsymbol{\alpha}^0\right)\right) \rightarrow \boldsymbol{\Sigma}_{{a}}
 $$ in probability with respect to the super-population model, where \newline $V\left\{S_{a, ih}\left(\boldsymbol{\eta}^0, \boldsymbol{\alpha}^0\right)\right\}$ is the variance of $S_{\boldsymbol{a},ih}(\boldsymbol{\eta}^0, \boldsymbol{\alpha}^0)$ with respect to the super-population model, and $S_{\boldsymbol{a},ih}(\boldsymbol{\eta}, \boldsymbol{\alpha})=\boldsymbol{a}^{\mathrm{T}} \boldsymbol{S}_{ih}\left(\boldsymbol{\eta}, \boldsymbol{\alpha}\right).$\label{Poi: CLT}
\item For $(\boldsymbol{\eta}, \boldsymbol{\alpha}) \in \mathcal{B}$, there exists a positive definitive and non-stochastic matrix $\boldsymbol{\Sigma}_{S}(\boldsymbol{\eta}, \boldsymbol{\alpha})$ such that $\bar{n}_0 \boldsymbol{V}\left\{\hat{\boldsymbol{S}}_{w}(\boldsymbol{\eta}, \boldsymbol{\alpha}) \mid\right.$ $\mathcal{F}\} \rightarrow \Sigma_{S}(\boldsymbol{\eta},\boldsymbol{\alpha})$ in probability with respect to the super-population model uniformly over $\mathcal{B}$, where $\bar{n}_0=\sum_{i=1}^m\bar{n}_i$, $\boldsymbol{x}^{\otimes 2}=\boldsymbol{x} \boldsymbol{x}^{\mathrm{T}}$ for a vector $\boldsymbol{x}$, and $$\boldsymbol{V}\left\{\hat{\boldsymbol{S}}_{w}(\boldsymbol{\eta}, \boldsymbol{\alpha}) \mid \mathcal{F}\right\}=\sum_{i=1}^{m}W_i^2N_{i}^{-2}\sum_{h=1}^{N_{i}} \pi_{ih}^{-1}\left(1-\pi_{ih}\right) \boldsymbol{S}_{ih}\left(\boldsymbol{\eta}, \boldsymbol{\alpha}\right)^{\otimes 2}$$ is the design variance of $\hat{\boldsymbol{S}}_{\omega}(\boldsymbol{\eta}, \boldsymbol{\alpha})$.\label{Poi: variance convergence}
\item $\sup _{\left(\boldsymbol{\eta}, \boldsymbol{\alpha} \right) \in \mathcal{B}} E\left\{\|{\boldsymbol{I}}_{i}\left(\boldsymbol{\eta}, \boldsymbol{\alpha} \right)\|^{2}\right\}<\infty$, and $\mathcal{I}\left(\boldsymbol{\eta}^{0}, \boldsymbol{\alpha}^{0}\right)$ is invertible, where \newline $\boldsymbol{\mathcal { I }} (\boldsymbol{\eta}, \boldsymbol{\alpha})$ is the limit of $\sum_{i=1}^{m}W_i\boldsymbol{\mathcal { I }}_{i}(\boldsymbol{\eta}, \boldsymbol{\alpha}) = \sum_{i=1}^{m}W_i E\{{\boldsymbol{I}}_{i}\left(\boldsymbol{\eta}, \boldsymbol{\alpha} \right)\}$ as $m\to\infty$.\label{Poi: I}
\item The number of clusters $K$, and the dimensions of covariates $p$ and $q$ are fixed. Conditions  (C\ref{cond:penalty})--(C\ref{cond:boundx}) hold.\label{Poi: Others}
 \end{enumerate}
Condition~(A\ref{Poi: true parameter}) is similar to Condition~(C\ref{cond:para}), and its second part is used to validate Condition~(C\ref{cond:var}). Condition~(A\ref{Poi: inc prob}) guarantees the central limit theorem by assuming independence between different elements in the finite population, and the independence is with respect to the model. Condition~(A\ref{Poi: exp ss}) regulates inclusion probabilities as well as population and sample sizes, and it is required to show the convergence rate in Condition~(C\ref{cond:clt}). Condition~(A\ref{Poi: estimating function}) is used to show the central limit theorem for the estimating function. Conditions~(A\ref{Poi: CLT})--(A\ref{Poi: variance convergence}) assume the probability limit for the scaled estimating function with respect to Poisson sampling conditional on the finite population, and it is also used to verify the central limit theorem for the estimating function. Condition~(A\ref{Poi: I}) is required to show the convergence of the information matrix, and  Condition~(A\ref{Poi: Others}) contains regularity conditions on the penalty term as well as others for the model.
 
 \subsubsection{Lemma 1}
 Under Poisson sampling and Conditions~(A\ref{Poi: true parameter})-(A\ref{Poi: exp ss}), Condition~(C\ref{cond:var}) holds.
\begin{proof}
 $$\begin{aligned}
 \boldsymbol{V}\{L_{\omega} \left(\boldsymbol{\eta, \alpha}\right)|\mathcal{F}\} &= \boldsymbol{V}\left\{\sum_{i=1}^{m}W_iN_{i}^{-1} \sum_{h=1}^{N_{i}}\delta_{ih}\pi_{ih}^{-1}L_{ih}\left(\boldsymbol{\eta}, \boldsymbol{\alpha} \right)|\mathcal{F}\right\}\\
 &= \sum_{i=1}^{m} W_i^2N_{i}^{-2} \sum_{h=1}^{N_i} \pi_{ih}^{-1} \left(1-\pi_{ih}\right)L_{ih}^{2}\left(\boldsymbol{\eta}, \boldsymbol{\alpha} \right)\\
 &\leq \sum_{i=1}^{m} W_i^2(C_1 \bar{n}_i)^{-1} {N_i}^{-1} \sum_{h=1}^{N_i}L_{ih}^{2}\left(\boldsymbol{\eta}, \boldsymbol{\alpha} \right)\\
 &=\sum_{i=1}^{m} W_i^2O_{p}({\bar{n}_i}^{-1}),
 \end{aligned}$$
 where the third inequality is based on Condition~(A\ref{Poi: exp ss}), and the last equality holds by Conditions~(A\ref{Poi: true parameter})-(A\ref{Poi: inc prob}) and weak law of large numbers.  Since  $\sum_{i=1}^mW_i=1$ and $W_i\asymp m^{-1}$ by Condition~(A\ref{Poi: exp ss}), $V\{L_{\omega} \left(\boldsymbol{\eta, \alpha}\right)|\mathcal{F}\} \rightarrow 0$ in probability with respect to the super-population model as ${n}_{i} \rightarrow \infty$ under Poisson sampling.
 
 
 
 This completes the proof of Lemma 1.
 \end{proof}
 
 \subsubsection{Lemma 2}
 Under Poisson sampling and Conditions(A\ref{Poi: inc prob})-(A\ref{Poi: estimating function}), we have
 $$
 \sup _{(\boldsymbol{\eta}, \boldsymbol{\alpha}) \in \mathcal{B}} \hat{V}\left\{\boldsymbol{a}^{\mathrm{T}} \hat{\boldsymbol{S}}_{\omega} (\boldsymbol{\eta}, \boldsymbol{\alpha}) \mid \mathcal{F}\right\} \asymp \bar{n}_{0}^{-1}
 $$
 in probability with respect to the super-population model and the sampling mechanism, where $$\hat{V}\left\{\boldsymbol{a}^{\mathrm{T}} \hat{\boldsymbol{S}}_{\omega} (\boldsymbol{\eta}, \boldsymbol{\alpha})  \mid \mathcal{F}\right\}= \sum_{i=1}^{m} W_i^2N_{i}^{-2} \sum_{h=1}^{N_i} \delta_{ih} {\pi_{ih}}^{-2}({1-\pi_{ih}}) \{\boldsymbol{a}^{\mathrm{T}}{\boldsymbol{S}}_{ih} (\boldsymbol{\eta}, \boldsymbol{\alpha})\}^{2},$$ where $\boldsymbol{a} \in \mathbb{R}^{q+ Kp}$ satisfying $\|\boldsymbol{a}\|=1$.
 
\begin{proof}
 First we consider $E\left[\hat{V}\left\{\boldsymbol{a}^{\mathrm{T}} \hat{\boldsymbol{S}}_{\omega} (\boldsymbol{\eta}, \boldsymbol{\alpha})  \mid \mathcal{F}\right\}\right]$ with respect to the
 super-population model and the sampling mechanism.
 \begin{align*}
&  E\left[\hat{V}\left\{\boldsymbol{a}^{\mathrm{T}} \hat{\boldsymbol{S}}_{\omega} (\boldsymbol{\eta}, \boldsymbol{\alpha}) \mid \mathcal{F}\right\}\right] \\
 & =E\left(E\left[\hat{V}\left\{\boldsymbol{a}^{\mathrm{T}} \hat{\boldsymbol{S}}_{\omega} (\boldsymbol{\eta}, \boldsymbol{\alpha}) \mid \mathcal{F}\right\} \mid \mathcal{F}\right]\right) \\
 & = E\left[ \sum_{i=1}^{m}W_i^2 N_{i}^{-2} \sum_{h=1}^{N_i} \delta_{ih} {\pi_{ih}}^{-2}({1-\pi_{ih}}) \boldsymbol{a}^{\mathrm{T}} {\boldsymbol{S}}_{ih} (\boldsymbol{\eta}, \boldsymbol{\alpha})^{2} \mid \mathcal{F}\right]\\
 &=  \sum_{i=1}^{m}W_i^2 N_{i}^{-2} \sum_{h=1}^{N_i} {\pi_{ih}}^{-1}({1-\pi_{ih}}) \boldsymbol{a}^{\mathrm{T}}{\boldsymbol{S}}_{ih} (\boldsymbol{\eta}, \boldsymbol{\alpha})^{2}.\\
 \end{align*}
 
 By Conditions (A\ref{Poi: exp ss})-(A\ref{Poi: estimating function}), we have
 $$ E\left[\hat{V}\left\{\boldsymbol{a}^{\mathrm{T}} \hat{\boldsymbol{S}}_{\omega} (\boldsymbol{\eta}, \boldsymbol{\alpha})  \mid \mathcal{F}\right\}\right] \asymp \bar{n}_{0}^{-1}$$
 uniformly over $\mathcal{B},$ where the expectation is taken with respect to super-population model.
 
 Next, consider the variance of $\hat{V}\left\{\boldsymbol{a}^{\mathrm{T}} \hat{\boldsymbol{S}}_{\omega} (\boldsymbol{\eta}, \boldsymbol{\alpha}) \mid \mathcal{F}\right\}$ with respect to the super-population model and the sampling mechanism:
 $$\begin{aligned}
 &V\left[\hat{V}\left\{\boldsymbol{a}^{\mathrm{T}}\hat{\boldsymbol{S}}_{\omega} (\boldsymbol{\eta}, \boldsymbol{\alpha}) \mid \mathcal{F}\right\}\right]\\
 &=E\left(V\left[\hat{V}\left\{\boldsymbol{a}^{\mathrm{T}} \hat{\boldsymbol{S}}_{\omega} (\boldsymbol{\eta}, \boldsymbol{\alpha}) \mid \mathcal{F}\right\} \mid \mathcal{F}\right]\right)\\&+V\left(E\left[\hat{V}\left\{\boldsymbol{a}^{\mathrm{T}} \hat{\boldsymbol{S}}_{\omega} (\boldsymbol{\eta}, \boldsymbol{\alpha}) \mid \mathcal{F}\right\} \mid \mathcal{F}\right]\right) .
 \end{aligned}
 $$
 
 Under Poisson sampling, we have 
 \begin{align*}
 & E\left(V\left[\hat{V}\left\{\boldsymbol{a}^{\mathrm{T}} \hat{\boldsymbol{S}}_{\omega} (\boldsymbol{\eta}, \boldsymbol{\alpha}) \mid \mathcal{F}\right\} \mid \mathcal{F}\right]\right)\\ 
 &= E\left[V\left\{\sum_{i=1}^{m} W_i^2N_{i}^{-2} \sum_{h=1}^{N_i} \delta_{ih} {\pi_{ih}}^{-2}({1-\pi_{ih}})\boldsymbol{a}^{\mathrm{T}} {\boldsymbol{S}}_{ih} (\boldsymbol{\eta}, \boldsymbol{\alpha})^{2} \mid \mathcal{F}\right\}\right]\\
 &= \sum_{i=1}^{m} E\left[V\left\{ W_i^2N_{i}^{-2} \sum_{h=1}^{N_i} \delta_{ih} {\pi_{ih}}^{-2}({1-\pi_{ih}}) \boldsymbol{a}^{\mathrm{T}} {\boldsymbol{S}}_{ih} (\boldsymbol{\eta}, \boldsymbol{\alpha})^{2} \mid \mathcal{F}\right\}\right]\\
 &= \sum_{i=1}^{m} W_i^4N_{i}^{-4} \sum_{h=1}^{N_i} \pi_{ih}(1-\pi_{ih}) {\pi_{ih}}^{-4}({1-\pi_{ih}})^2 E\left\{\boldsymbol{a}^{\mathrm{T}}{\boldsymbol{S}}_{ih} (\boldsymbol{\eta}, \boldsymbol{\alpha})^{4}\right\}\\
 &= \sum_{i=1}^{m} W_i^4N_{i}^{-4} \sum_{h=1}^{N_i} {\pi_{ih}}^{-3}({1-\pi_{ih}})^3 E\left\{\boldsymbol{a}^{\mathrm{T}}{\boldsymbol{S}}_{ih} (\boldsymbol{\eta}, \boldsymbol{\alpha})^{4}\right\}\\
 & \leq \sum_{i=1}^{m} W_i^4(C_1\bar{n}_{i})^{-3}N_{i}^{-1} \sum_{h=1}^{N_i} E\left\{\boldsymbol{a}^{\mathrm{T}} {\boldsymbol{S}}_{ih} (\boldsymbol{\eta}, \boldsymbol{\alpha})^{4}\right\}\\
 & \asymp \bar{n}_{0}^{-3},
 \end{align*}
 
 and 
\begin{align*}
 & V\left(E\left[\hat{V}\left\{\boldsymbol{a}^{\mathrm{T}} \hat{\boldsymbol{S}}_{\omega} (\boldsymbol{\eta}, \boldsymbol{\alpha}) \mid \mathcal{F}\right\} \mid \mathcal{F}\right]\right)\\
 = & V \left\{ \sum_{i=1}^{m}W_i^2 N_{i}^{-2} \sum_{h=1}^{N_i} {\pi_{ih}}^{-1}({1-\pi_{ih}}) \boldsymbol{a}^{\mathrm{T}}{\boldsymbol{S}}_{ih} (\boldsymbol{\eta}, \boldsymbol{\alpha})^{2}\right\}\\
 =&  \sum_{i=1}^{m}W_i^4N_{i}^{-4} \sum_{h=1}^{N_i} V \left\{  {\pi_{ih}}^{-1}({1-\pi_{ih}}) \boldsymbol{a}^{\mathrm{T}}{\boldsymbol{S}}_{ih} (\boldsymbol{\eta}, \boldsymbol{\alpha})^{2}\right\}\\
 \leq & \sum_{i=1}^{m}W_i^4N_{i}^{-4} \sum_{h=1}^{N_i} E \left[ {\pi_{ih}}^{-2}({1-\pi_{ih}})^{2} \left\{ \boldsymbol{a}^{\mathrm{T}}{\boldsymbol{S}}_{ih} (\boldsymbol{\eta}, \boldsymbol{\alpha})\right\}^{4} \right]\\
  \leq& \sum_{i=1}^{m} W_i^4(C_1 \bar{n}_{i})^{-2}N_{i}^{-2} \sum_{h=1}^{N_i} E \left[ ({1-\pi_{ih}})^{2} \left\{ \boldsymbol{a}^{\mathrm{T}}{\boldsymbol{S}}_{ih} (\boldsymbol{\eta}, \boldsymbol{\alpha})\right\}^{4} \right]\\
  =& o(1)
 \end{align*}
 uniformly over $\mathcal{B}$ by Condition~(A\ref{Poi: estimating function}). This completes the proof of Lemma 2.
 \end{proof}
  \subsubsection{Lemma 3}
  Under Poisson sampling and Conditions (A\ref{Poi: inc prob})-(A\ref{Poi: variance convergence}), Condition (C\ref{cond:clt}) holds.
\begin{proof}
  Given every $\boldsymbol{a} \in \mathbb{R}^{q+ Kp}$ satisfying $\|\boldsymbol{a}\|=1$, by Conditions (A\ref{Poi: inc prob})-(A\ref{Poi: CLT}), Theorem 1.3.5 of \cite{fuller2011sampling} shows 
  \[ \hat{V}\left\{\boldsymbol{a}^{\mathrm{T}}\hat{\boldsymbol{S}}_{\omega} (\boldsymbol{\eta}^{0}, \boldsymbol{\alpha}^{0}) \mid \mathcal{F}\right\}^{-1/2}\boldsymbol{a}^{\mathrm{T}}\left\{\hat{\boldsymbol{S}}_{\omega} (\boldsymbol{\eta}^{0}, \boldsymbol{\alpha}^{0}) - {\boldsymbol{S}}_{\omega} (\boldsymbol{\eta}^{0}, \boldsymbol{\alpha}^{0})\right\} \mid  \rightarrow N(0,1)
  \]
 in distribution with respect to the sampling mechanism conditional on finite population in
 probability, where $\hat{\boldsymbol{S}}_{\omega}(\boldsymbol{\eta}^{0}, \boldsymbol{\alpha}^{0})= \sum_{i=1}^{m}W_iN_{i}^{-1} \sum_{h=1}^{N_{i}}\delta_{ih}\pi_{ih}^{-1}\boldsymbol{S}_{ih}\left(\boldsymbol{\eta}^{0}, \boldsymbol{\alpha}^{0}\right)$,   ${\boldsymbol{S}}_{\omega}(\boldsymbol{\eta}^{0}, \boldsymbol{\alpha}^{0})= E\{\hat{\boldsymbol{S}}_{\omega}(\boldsymbol{\eta}^{0}, \boldsymbol{\alpha}^{0})\mid \mathcal{F}\}.$
 
 Since the finite population  is a random sample generated from a super-population model satisfying Conditions (A\ref{Poi: estimating function})-(A\ref{Poi: CLT}), by the Chebychev’s inequality \citep[ Corollary 3.1.3]{athreya2006measure}, we have
 \[\boldsymbol{a}^{\mathrm{T}}{\boldsymbol{S}}_{\omega} (\boldsymbol{\eta}^{0}, \boldsymbol{\alpha}^{0}) = \boldsymbol{a}^{\mathrm{T}}\left[{\boldsymbol{S}}_{\omega} (\boldsymbol{\eta}^{0}, \boldsymbol{\alpha}^{0}) - E\left\{{\boldsymbol{S}}_{\omega} (\boldsymbol{\eta}^{0}, \boldsymbol{\alpha}^{0})\right\}\right] = o_p(\bar{n}_0^{-1/2})\]
 with respect to the super-population model, where the second equality holds since $E\left\{{\boldsymbol{S}}_{\omega} (\boldsymbol{\eta}^{0}, \boldsymbol{\alpha}^{0})\right\} = 0$, and the last equality holds by Condition (A\ref{Poi: exp ss}).
 
 Thus, by Condition (A\ref{Poi: variance convergence}), Lemma 3 and Theorem 5.1 of \cite{rubin2005two} we have 
 \[ \hat{V}\left\{\boldsymbol{a}^{\mathrm{T}}\hat{\boldsymbol{S}}_{\omega} (\boldsymbol{\eta}^{0}, \boldsymbol{\alpha}^{0}) \mid \mathcal{F}\right\}^{-1/2}\boldsymbol{a}^{\mathrm{T}}\hat{\boldsymbol{S}}_{\omega} (\boldsymbol{\eta}^{0}, \boldsymbol{\alpha}^{0})  \rightarrow N(0,1)
  \]
 in distribution with respect to the super-population model and the sampling mechanism,
 which validates Condition (C\ref{cond:clt}) by the Cramér-Wold device.
\end{proof}
  
  \subsubsection{Lemma 4}
  Under Poisson sampling, Conditions (A\ref{Poi: inc prob})-(A\ref{Poi: exp ss}) and Condition (A\ref{Poi: I}), Condition (C\ref{cond:infmat}) holds.
  
\begin{proof}
  Consider
  \[\begin{aligned}
  & E\left\{\hat{\boldsymbol{I}}_{\omega}(\boldsymbol{\eta, \alpha})  \mid \mathcal{F}\right\} = \sum_{i=1}^{m}W_iN_{i}^{-1} \sum_{h=1}^{N_{i}}\boldsymbol{I}_{ih}\left(\boldsymbol{\eta}, \boldsymbol{\alpha}\right)\\
  & E\left\{\sum_{i=1}^{m}{\boldsymbol{I}}_{i}\left(\boldsymbol{\eta}, \boldsymbol{\alpha}\right) \right\}  =\boldsymbol{\mathcal { I }} (\boldsymbol{\eta}, \boldsymbol{\alpha}).
  \end{aligned}\]
  
  Then we have 
  \[\begin{aligned} E\left\{\hat{\boldsymbol{I}}_{\omega}(\boldsymbol{\eta, \alpha})\right\} & = E\left[E\left\{\hat{\boldsymbol{I}}_{\omega}(\boldsymbol{\eta, \alpha})  \mid \mathcal{F}\right\} \right]\\
  & = \boldsymbol{\mathcal { I }} (\boldsymbol{\eta}, \boldsymbol{\alpha})
  \end{aligned}
  \]
  with respect to the super-population model and the sampling mechanism. For $\boldsymbol{a} \in \mathbb{R}^{q+ Kp}$ satisfying $\|\boldsymbol{a}\|=1$, consider the variance of $\boldsymbol{a}^{\mathrm{T}} \hat{\boldsymbol{I}}_{\omega}(\boldsymbol{\eta, \alpha})  \boldsymbol{a}$ with respect to the super-population
  model and the sampling mechanism:
  \[V\left\{\boldsymbol{a}^{\mathrm{T}} \hat{\boldsymbol{I}}_{\omega}(\boldsymbol{\eta, \alpha})  \boldsymbol{a}\right\}=E\left[V\left\{\boldsymbol{a}^{\mathrm{T}} \hat{\boldsymbol{I}}_{\omega}(\boldsymbol{\eta, \alpha})  \boldsymbol{a} \mid \mathcal{F}\right\}\right]+V\left[E\left\{\boldsymbol{a}^{\mathrm{T}} \hat{\boldsymbol{I}}_{\omega}(\boldsymbol{\eta, \alpha}) \boldsymbol{a} \mid \mathcal{F}\right\}\right].\]
  
  First consider
  \[\begin{aligned}
  V\left\{\boldsymbol{a}^{\mathrm{T}} \hat{\boldsymbol{I}}_{\omega}(\boldsymbol{\eta, \alpha})  \boldsymbol{a} \mid \mathcal{F}\right\} &=\sum_{i=1}^{m}W_i^2{N_{i}^{-2}} \sum_{h=1}^{N_i} \frac{1-\pi_{ih}}{\pi_{ih}} Z_{\boldsymbol{a}, ih}^{2}(\boldsymbol{\eta, \alpha}) \\
  & \leq \sum_{i=1}^{m} W_i^2{(C_{1} n_{i})^{-1}} \frac{1}{N_i} \sum_{h=1}^{N_i} Z_{\boldsymbol{a}, ih}^{2}(\boldsymbol{\eta, \alpha}),
  \end{aligned}
  \]
 where $Z_{\boldsymbol{a}, ih} = \boldsymbol{a}^{\mathrm{T}} \boldsymbol{I}_{ih}\left(\boldsymbol{\eta}, \boldsymbol{\alpha}\right) \boldsymbol{a},$ the second inequality holds by Conditions (A\ref{Poi: inc prob})-(A\ref{Poi: exp ss}). Thus we have 
 \[E\left[V\left\{\boldsymbol{a}^{\mathrm{T}} \hat{\boldsymbol{I}}_{\omega}(\boldsymbol{\eta, \alpha})  \boldsymbol{a} \mid \mathcal{F}\right\}\right] = O(\bar{n}_{0}^{-1})
 \]
 in probability with respect to the super-population model uniformly over $\mathcal{B}$ by Condition(A\ref{Poi: I}).
 
 Next consider
 \[\begin{aligned}
 V\left\{\sum_{i=1}^{m}W_iN_{i}^{-1} \sum_{h=1}^{N_{i}} \boldsymbol{a}^{\mathrm{T}} \boldsymbol{I}_{ih}\left(\boldsymbol{\eta}, \boldsymbol{\alpha}\right) \boldsymbol{a} \right\}& = \sum_{i=1}^{m}W_i^2N_{i}^{-1} V \left\{ \boldsymbol{a}^{\mathrm{T}} \boldsymbol{I}_{ih}\left(\boldsymbol{\eta}, \boldsymbol{\alpha}\right) \boldsymbol{a} \right\}\\
 & \leq \sum_{i=1}^{m}W_i^2N_{i}^{-1} E\left\{|\boldsymbol{a}^{\mathrm{T}} \boldsymbol{I}_{ih}\left(\boldsymbol{\eta}, \boldsymbol{\alpha}\right) \boldsymbol{a}|^2\right\}\\
 &=o(1)
 \end{aligned}
 \]
 in probability with respect to the super-population model uniformly for $(\boldsymbol{\eta}, \boldsymbol{\alpha}) \in \mathcal{B}$ by Condition (A\ref{Poi: I}), where the last equality holds by Condition (A\ref{Poi: exp ss}).
 
 By Markov’s inequality, we can show that
 \[\hat{\boldsymbol{I}}_{\omega}(\boldsymbol{\eta, \alpha})  \rightarrow \boldsymbol{\mathcal { I }} (\boldsymbol{\eta}, \boldsymbol{\alpha}) 
 \]
 in probability with respect to the super-population model and the sampling mechanism
 uniformly for $(\boldsymbol{\eta}, \boldsymbol{\alpha}) \in \mathcal{B}$, which completes the proof of Lemma 4.
 \end{proof}

\subsection{Stratified multistage cluster sampling}\label{app: multistage cluster sampling}
In this section, we consider a popular stratified multistage cluster sampling as in Example~1 of \cite{kim2023}; see Example~1 of \cite{Rubin_Kratina} for a similar treatment.  We assume that elements within each stratum share  common  regression coefficients, and we estimate the cluster structure by  regression coefficients on the stratum level. 

Assume that there exist $m$ strata in the finite population $\mathcal{F}$. Let $N_{i}$ be the number of clusters in the $i$th stratum, $M_{ih}$ be the number of elements in the $h$th cluster of the $i$th stratum, and $M_i=\sum_{h=1}^{N_i}M_{ih}$ be the  size of the $i$th stratum, where $N_i$ is the number of clusters in the $i$th stratum.  Let $n_i$ be the number of clusters selected from the $i$th stratum  with selection probability $p_{ih}$ satisfying $\sum_{h=1}^{N_i}p_{ih}=1$ for $i=1,\ldots,m$, where $p_{ih}$ is proportional to a size measure, for example, $p_{ih}\propto M_{ih}$ for $h=1,\ldots,N_i$. Within each selected cluster, denote $\hat{L}_{ih}(\boldsymbol{\eta},\boldsymbol{\beta})$ and $\hat{S}_{w,ih}(\boldsymbol{\eta},\boldsymbol{\beta})$ as  design-unbiased estimators of the associated cluster means of the objective function and score function under sampling at the second and subsequent stages. Then, the probability-weighted score function is $\hat{S}_w(\boldsymbol{\eta},\boldsymbol{\beta}) = \sum_{i=1}^mW_i\hat{S}_{w,i}(\boldsymbol{\eta},\boldsymbol{\beta})$, where $W_i \propto M_i$ and $\sum_{i=1}^mW_i=1$, $\hat{S}_{w,i}(\boldsymbol{\eta},\boldsymbol{\beta}) = n_i^{-1}\sum_{h=1}^{n_i}\tilde{S}_{w,ih}(\boldsymbol{\eta},\boldsymbol{\beta})$, $\tilde{S}_{w,ih}(\boldsymbol{\eta},\boldsymbol{\beta}) = d_{a(ih)}p^{-1}{a(ih)}\hat{S}_{w,a(ih)}(\boldsymbol{\eta},\boldsymbol{\beta})$, $d_{a(ih)}=M_{a(ih)}/M_{i}$, $a(ih)$ is the index of the $h$th selected cluster. We implicitly assume that $m\to\infty$, and the number of stages are fixed for the sampling design.  For this sampling design, we make the following assumptions.

\begin{enumerate}
\renewcommand{\labelenumi}{(B\arabic{enumi})}
    \item There exists constants $0<C_1<C_2<\infty$, such that $C_1<M_{i}p_{ih}<C_2$ almost surely with respect to the model for $i=1,\ldots,m$ and $h=1,\ldots,N_i$.\label{SRS: SELEC}
    \item $\max_{1\leq i\leq m}n_i=O(1)$.\label{SRS: SSsize}
    \item There exists $\delta>0$ such that $$\sum_{i=1}^mW_iE\left\{\lVert \tilde{S}_{w,a(i1)}(\boldsymbol{\eta},\boldsymbol{\alpha}) - {S}_{i}(\boldsymbol{\eta},\boldsymbol{\alpha})\rVert^{2+\delta} \mid \mathcal{F}\right\}=O_p(1)$$ with respect to the model uniformly for $(\boldsymbol{\eta},\boldsymbol{\alpha})\in\mathcal{B}$, where  $\mathcal{B}$ is a compact set containing $(\boldsymbol{\eta}^0,\boldsymbol{\alpha}^0)$ as an interior point, and ${S}_{i}(\boldsymbol{\eta},\boldsymbol{\alpha})=E\{\tilde{S}_{w,a(i1)}(\boldsymbol{\eta},\boldsymbol{\alpha}) \mid\mathcal{F}\}$ is the average score function associated with the $i$th stratum.\label{SRS: clt}
    \item $\max_{1\leq i\leq m}\sup_{(\boldsymbol{\eta},\boldsymbol{\alpha})\in\mathcal{B}}E\left\{\lVert \hat{I}_{w,a(i1)}(\boldsymbol{\eta},\boldsymbol{\alpha})\rVert^2\mid \mathcal{F}\right\}=O_p(1)$ with respect to the model, where $\hat{I}_{w,ih}(\boldsymbol{\eta},\boldsymbol{\alpha})=\partial\hat{S}_{w,ih}(\boldsymbol{\eta},\boldsymbol{\alpha})/\partial(\boldsymbol{\eta},\boldsymbol{\alpha}).$\label{SRS: I conv}
    \item There exists a non-stochastic function $\mathcal{I}(\boldsymbol{\eta},\boldsymbol{\alpha})$ such that 
    $$\sup_{(\boldsymbol{\eta},\boldsymbol{\alpha})\in\mathcal{B}}\lVert I(\boldsymbol{\eta},\boldsymbol{\alpha}) - \mathcal{I}(\boldsymbol{\eta},\boldsymbol{\alpha})\rVert\to0$$ in probability with respect to the model, where 
    \begin{align*}
         & I(\boldsymbol{\eta},\boldsymbol{\alpha})=E\{\hat{I}_w(\boldsymbol{\eta},\boldsymbol{\alpha})\mid\mathcal{F}\},\,\hat{I}_w(\boldsymbol{\eta},\boldsymbol{\alpha}) = \sum_{i=1}^mW_i\hat{I}_{w,i}(\boldsymbol{\eta},\boldsymbol{\beta})\\
         &\hat{I}_{w,i}(\boldsymbol{\eta},\boldsymbol{\beta}) = n_i^{-1}\sum_{h=1}^{n_i}\tilde{I}_{w,ih}(\boldsymbol{\eta},\boldsymbol{\beta}),\, \tilde{I}_{w,ih}(\boldsymbol{\eta},\boldsymbol{\beta}) = d_{a(ih)}p^{-1}{a(ih)}\hat{I}_{w,a(ih)}(\boldsymbol{\eta},\boldsymbol{\beta}),
    \end{align*}
and $\mathcal{I}(\boldsymbol{\eta}^0,\boldsymbol{\alpha}^0)$ is invertible.\label{SRS: I CONVERGENCE}
    \item $\sup_{(\boldsymbol{\eta},\boldsymbol{\alpha})\in\mathcal{B}}\lVert n_0\sum_{i=1}^mW_h^2V\{\tilde{S}_{w,a(i1)}(\boldsymbol{\eta},\boldsymbol{\alpha})\mid\mathcal{F}\}-\Sigma_S(\boldsymbol{\eta},\boldsymbol{\alpha})\rVert\to0$ in probability with respect to the model, where $\Sigma_S(\boldsymbol{\eta},\boldsymbol{\alpha})$ is a non-stochastic and positive definitive variance matrix for $(\boldsymbol{\eta},\boldsymbol{\alpha})\in\mathcal{B}$, and $n_0=\sum_{i=1}^mn_i$.\label{SRS: Var convergence}
    \item $\sup_{(\boldsymbol{\eta},\boldsymbol{\alpha})\in\mathcal{B}}V\{S(\boldsymbol{\eta},\boldsymbol{\alpha})\}=o(n_0^{-1})$ with respect to the model, where $S(\boldsymbol{\eta},\boldsymbol{\alpha})=E\{\hat{S}_w(\boldsymbol{\eta},\boldsymbol{\alpha})\mid \mathcal{F}\}$. \label{SRS: VAR rate}
      \item Conditions (C\ref{cond:size})--(C\ref{cond:var}), (C\ref{cond:inclusion})--(C\ref{cond:boundx}) hold. \label{SRS: Others}
\end{enumerate}
 Conditions (B\ref{SRS: SELEC})--(B\ref{SRS: SSsize}) are commonly assumed for stratified multistage cluster sampling \cite{krewski1981inference}. That is, we consider a sampling design, where the number of selected clusters is bounded but the number of strata diverges. Condition (B\ref{SRS: clt}) is used to show the central limit theorem in Condition (C\ref{cond:clt}), and Conditions (B\ref{SRS: I conv})--(B\ref{SRS: I CONVERGENCE}) are mainly used to show the convergence result in Condition (C\ref{cond:infmat}). Conditions (B\ref{SRS: Var convergence})--(B\ref{SRS: VAR rate}) shows the convergence of the sample variance conditional on the finite population, and the variance $V\{S(\boldsymbol{\eta},\boldsymbol{\beta})\}$ is asymptotically negligible, which is required by the second part of Condition (C\ref{cond:penalty}). Condition (B\ref{SRS: Others}) regulates the objective function as well as the penalty, and the variance of the probability-weighted objective function is also assumed to converge in probability. Specifically, by Condition~(B\ref{SRS: Others}), we can show  $\max_{1\leq i\leq m}W_i = O_p(m^{-1})$, which is essential to establish convergence rates under stratified multi-stage cluster sampling.

Different from Poisson sampling, the sampling indicators may be correlated due to sampling at the second and subsequent stages. Under the above conditions, Conditions~(C\ref{cond:clt})--(C\ref{cond:infmat}) can be verified in exactly the same manner as Lemmas~S5--S6 of \cite{kim2023}, so the detailed proofs are omitted.

\section{More simulation results}
\label{sec:more_examples}

In this section, we present the results of our proposed estimator when the data sets are simulated from settings that are different from the assumed models. 

\subsection{Example 1}

\begin{figure}[h]
    \centering
    \includegraphics[scale=0.6]{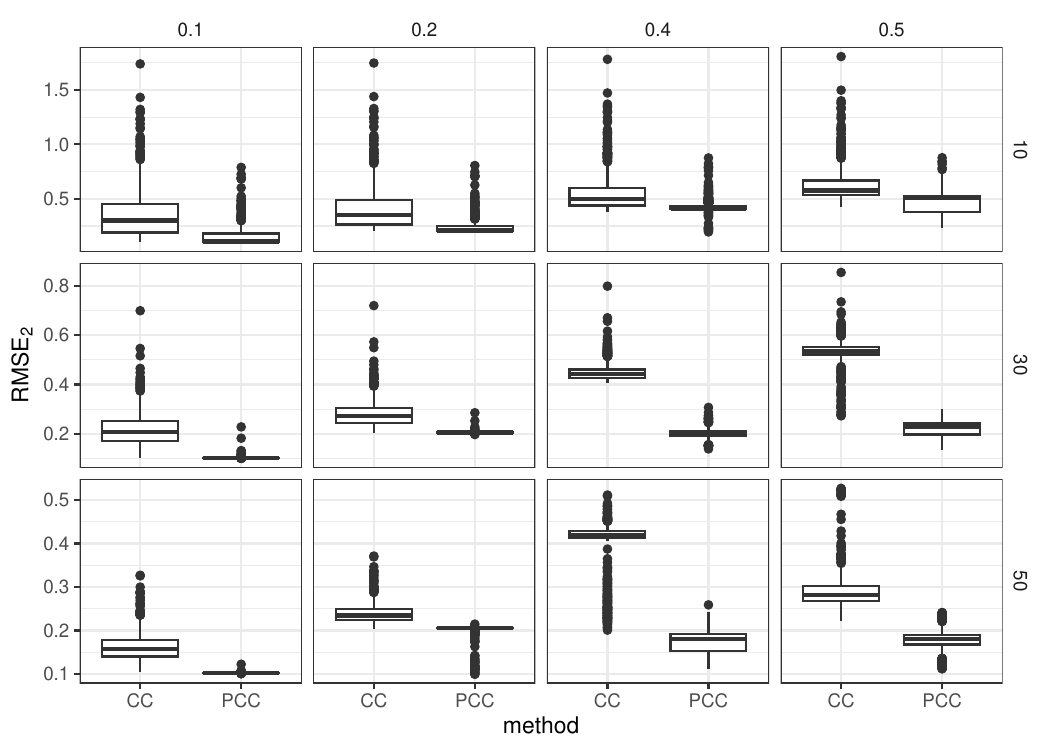}
    \caption{$\text{RMSE}_2$ based on 1\,000 simulations with expected sample size $ n_i = 10$ (top), $n_i = 30$ (middle) and $n_i=50$ (bottom).}
    \label{fig:lmm_rmse2}
\end{figure}

In this example, we generate finite populations from linear mixed models, that is 
\begin{equation}
\label{eq:more1}
 y_{ih}  = u_i +  \beta_{0,i} + \beta_{1,i} x_{ih} + \epsilon_{ih},   
\end{equation}
where $u_i$ is a random effect with $u_i \overset{iid}{\sim} N(0, \epsilon_u^2)$. This model used a lot in small area estimation in survey sampling, see examples in \cite{rao2015small}. Other settings are the same as those in Section \ref{subsec:sim_linear}. Note that,  \cite{liu2021capturing} did a similar comparison without employing a survey sampling setup. In their comparison, they simulated datasets from a linear mixed model with common regression coefficients and explored the performance of the clustered intercept model for capturing the random effects. In our model, the values of $\beta_{0,i}'s$ can contain the information of random effects. We explore and compare the results for different values of $\epsilon_u = 0.1, 0.2, 0.4, 0.5$ for CC and PCC.  Table \ref{table_lmm} shows the estimated $\hat{K}$ and ARI for different setups. It can be seen that when $\epsilon_{u}$ and $n_i$ are larger, the estimated $\hat{K}$ is much larger than the true number of clusters. This is because that the estimated values of $\beta_{0,i}'s$  have the information of $u_i$. In Figure \ref{fig:lmm_rmse2}, we calculate RMSEs defined as $\text{RMSE}_2 = \frac{1}{m}\sum_{i=1}^m (\beta_{0,i} + u_i - \hat{\beta}_{0,i})^2$, which measures the accuracy of using fixed clustered effects to estimate the fixed intercept and random effects together. From these figures, we observe that when $n_i = 30, 50$ and $\epsilon_u = 0.4, 0.5$, PCC has much smaller RMSE for estimating $\beta_{i,0} + u_i$. These are consistent with the estimated number of clusters in Table \ref{table_lmm}, which implies that the $\hat{\beta}_{i,0}$ has information of $u_i$. CC can capture this information only when $\epsilon_u = 0.5$ and $n_i = 50$.

\begin{table}[th]
\centering
\caption{Summary of average $\hat{K}$ and average ARI of CC and PCC for the linear mixed model based on 1\,000 simulations, along with standard deviations in the parentheses. }
\label{table_lmm}
\begin{tabular}{lllllll}
  \hline &  &  \multicolumn{2}{c}{$\hat{K}$} & \multicolumn{2}{c}{ ARI } \\
& &  CC & PCC & CC & PCC \\
  \hline
 \multirow{3}{*}{$\epsilon_u=0.1$} & $n_i = 10$ &5.01(1.089) & 3.41(0.578) & 0.94(0.042) & 0.99(0.019) \\ 
 & $n_i = 30$  &4.94(0.739) & 3.05(0.232) & 0.95(0.042) & 1(0.01) \\ 
 & $n_i = 50$  & 4.96(0.617) & 3.04(0.206) & 0.96(0.036) & 1(0.005) \\ 
\hline
  \multirow{3}{*}{$\epsilon_u=0.2$}  &  $n_i = 10$  & 5.00(1.085) & 3.41(0.587) & 0.94(0.04) & 0.99(0.019) \\ 
 & $n_i = 30$  & 4.85(0.66) & 3.04(0.204) & 0.96(0.033) & 1(0.006) \\ 
 &  $n_i = 50$  & 4.72(0.542) & 3.21(0.976) & 0.97(0.021) & 0.99(0.061) \\ 
    \hline
   \multirow{3}{*}{$\epsilon_u=0.4$}  & $n_i = 10$  &  5.03(1.088) & 3.54(0.932) & 0.93(0.042) & 0.98(0.049) \\ 
 & $n_i = 30$  & 4.70(0.527) & 10.4(1.288) & 0.97(0.017) & 0.56(0.051) \\ 
 & $n_i = 50$  &  5.30(2.468) & 11.06(1.668) & 0.92(0.142) & 0.5(0.064) \\ 
    \hline
  \multirow{3}{*}{$\epsilon_u=0.5$}  & $n_i = 10$  & 5.14(1.134) & 5.47(3.153) & 0.92(0.049) & 0.86(0.18) \\ 
 & $n_i = 30$  & 5.00(1.299) & 10.82(1.611) & 0.95(0.076) & 0.51(0.064) \\ 
&  $n_i = 50$  & 12.18(1.511) & 12.61(1.585) & 0.56(0.069) & 0.43(0.056) \\ 
   \hline
\end{tabular}
\end{table}

\subsection{Example 2}
In this example, we generate a finite population from the following model. 
\begin{equation}
\label{eq:more2}
  y_{ih} = \beta_{0,i} +\beta_{1,i} x_{ih} + b\log(\vert x_{ih}\vert ) + \epsilon_{ih}.
\end{equation}

Table \ref{tab:more2} summarizes the results of average $\hat{K}$ and average ARI based on 1\,000 simulations when $b = 0.3$ and 1. When $b=0.3$, the fitted model is not really different from the true model. PCC can identify the cluster structure well and performs better than CC.  When $b=1$, both approaches have worse performance compared to the case in Section \ref{subsec:sim_linear} when the sample size is small ($n_i=10$). The performances are similar to the case in Section \ref{subsec:sim_linear} when the sample size is large.

\begin{table}[ht]
\centering
\caption{Summary of average $\hat{K}$ and average ARI of CC and PCC for model \eqref{eq:more2} based on 1\,000 simulations, along with standard deviations in the parentheses. }
\label{tab:more2}
\begin{tabular}{lllllll}
  \hline &  &  \multicolumn{2}{c}{$\hat{K}$} & \multicolumn{2}{c}{ ARI } \\
& &  CC & PCC & CC & PCC \\
  \hline
\multirow{3}{*}{$b=0.3$} &  $n_i=10$&  4.86(1.034) & 3.41(0.565) & 0.94(0.039) & 0.98(0.024) \\ 
 &$n_i=30$ &  4.86(0.771) & 3.06(0.234) & 0.95(0.043) & 1(0.015) \\ 
 &$n_i=50$ &  4.89(0.637) & 3.05(0.233) & 0.96(0.037) & 1(0.013) \\ 
  \hline
  \multirow{3}{*}{$b=1$} &$n_i=10$ & 5.83(1.379) & 4.87(0.975) & 0.84(0.096) & 0.86(0.065) \\ 
 & $n_i=30$ &  4.51(0.590) & 3.17(0.378) & 0.96(0.022) & 0.99(0.024) \\ 
 & $n_i=50$ & 3.98(0.333) & 3.05(0.221) & 0.98(0.011) & 1(0.019) \\ 
   \hline
\end{tabular}
\end{table}

\subsection{Example 3}

In this example, we generate a finite population from the following model. In this example, the cluster structure is applied for a nonlinear term of $x_{ih}$.
\begin{equation}
\label{eq:more3}
   y_{ih} = \beta_{0,i} +\beta_{1,i} (x_{ih} +\log(\vert x_{ih}\vert) + \epsilon_{ih}. 
\end{equation}

Table \ref{tab:more3} summarizes the results of average $\hat{K}$ and average ARI. We observe that CC has a very small ARI when $n_i=10$, which indicates that it cannot recover the cluster structure. Our proposed PCC tends to estimate the number of clusters larger than the true value, but the ARI is still relatively large. 

\begin{table}[ht]
\centering
\caption{Summary of average $\hat{K}$ and average ARI of CC and PCC for model \eqref{eq:more3} based on 1\,000 simulations, along with standard deviations in the parentheses.\label{tab:more3} }
\begin{tabular}{lllll}
  \hline  &  \multicolumn{2}{c}{$\hat{K}$} & \multicolumn{2}{c}{ ARI } \\
 &  CC & PCC & CC & PCC \\
  \hline
$n_i=10$ &  3.98(2.264) & 4.73(1.054) & 0.20(0.265) & 0.91(0.044) \\ 
  $n_i=30$ &  5.54(0.787) & 3.79(0.694) & 0.92(0.042) & 0.97(0.021) \\ 
  $n_i=50$ & 4.86(0.451) & 4.11(0.687) & 0.96(0.015) & 0.96(0.019) \\ 
   \hline
\end{tabular}
\end{table}


\end{appendices}

\bibliographystyle{apalike}
\bibliography{reference}

\begin{thebibliography}{}

\bibitem[Amemiya, 1985]{takeshi1985advanced}
Amemiya, T. (1985).
\newblock {\em Advanced Econometrics}.
\newblock Harvard University Press, Cambridge.

\bibitem[Athreya and Lahiri, 2006]{athreya2006measure}
Athreya, K.~B. and Lahiri, S.~N. (2006).
\newblock {\em Measure Theory and Probability Theory}.
\newblock Springer, New York.

\bibitem[Azka~Ubaidillah and Mangku, 2019]{Ubaidilah2019}
Azka~Ubaidillah, Khairil Anwar~Notodiputro, A.~K. and Mangku, I.~W. (2019).
\newblock Multivariate fay-herriot models for small area estimation with
  application to household consumption per capita expenditure in indonesia.
\newblock {\em Journal of Applied Statistics}, 46(15):2845--2861.

\bibitem[Berg and Fuller, 2014]{berg2014small}
Berg, E.~J. and Fuller, W.~A. (2014).
\newblock Small area prediction of proportions with applications to the
  canadian labour force survey.
\newblock {\em Journal of Survey Statistics and Methodology}, 2(3):227--256.

\bibitem[Boyd et~al., 2011]{boyd2011distributed}
Boyd, S., Parikh, N., Chu, E., Peleato, B., Eckstein, J., et~al. (2011).
\newblock Distributed optimization and statistical learning via the alternating
  direction method of multipliers.
\newblock {\em Foundations and Trends in Machine learning}, 3(1):1--122.

\bibitem[Chen et~al., 2019]{chen2019subgroup}
Chen, K., Huang, R., Chan, N.~H., and Yau, C.~Y. (2019).
\newblock Subgroup analysis of zero-inflated poisson regression model with
  applications to insurance data.
\newblock {\em Insurance: Mathematics and Economics}, 86:8--18.

\bibitem[Datta and Lahiri, 2000]{datta2000unified}
Datta, G.~S. and Lahiri, P. (2000).
\newblock A unified measure of uncertainty of estimated best linear unbiased
  predictors in small area estimation problems.
\newblock {\em Statistica Sinica}, pages 613--627.

\bibitem[Dumitrescu et~al., 2021]{dumitrescu2021variable}
Dumitrescu, L., Qian, W., and Rao, J. (2021).
\newblock Variable selection for longitudinal survey data.
\newblock {\em arXiv preprint arXiv:2105.00504}.

\bibitem[Esteban et~al., 2020]{esteban2020small}
Esteban, M.~D., Lombard{\'\i}a, M.~J., L{\'o}pez-Vizca{\'\i}no, E., Morales,
  D., and P{\'e}rez, A. (2020).
\newblock Small area estimation of proportions under area-level compositional
  mixed models.
\newblock {\em Test}, 29:793--818.

\bibitem[Esteban et~al., 2022]{Esteban2022}
Esteban, M.~D., Lombardía, M.~J., López-Vizcaíno, E., Morales, D., and
  Pérez, A. (2022).
\newblock Empirical best prediction of small area bivariate parameters.
\newblock {\em Scandinavian Journal of Statistics}, 49(4):1699--1727.

\bibitem[Fan and Li, 2001]{fan2001variable}
Fan, J. and Li, R. (2001).
\newblock Variable selection via nonconcave penalized likelihood and its oracle
  properties.
\newblock {\em Journal of the American Statistical Association},
  96(456):1348--1360.

\bibitem[Fan and Lv, 2011]{fan2011nonconcave}
Fan, J. and Lv, J. (2011).
\newblock Nonconcave penalized likelihood with np-dimensionality.
\newblock {\em IEEE Transactions on Information Theory}, 57(8):5467--5484.

\bibitem[Fuller, 2011]{fuller2011sampling}
Fuller, W.~A. (2011).
\newblock {\em Sampling Statistics}.
\newblock Wiley, New Jersey.

\bibitem[Horvitz and Thompson, 1952]{horvitz1952generalization}
Horvitz, D.~G. and Thompson, D.~J. (1952).
\newblock A generalization of sampling without replacement from a finite
  universe.
\newblock {\em Journal of the American Statistical Association},
  47(260):663--685.

\bibitem[Hu et~al., 2021]{hu2021subgroup}
Hu, X., Huang, J., Liu, L., Sun, D., and Zhao, X. (2021).
\newblock Subgroup analysis in the heterogeneous {Cox} model.
\newblock {\em Statistics in Medicine}, 40(3):739--757.

\bibitem[Innocent~Ngaruye and Singull, 2017]{Ngaruye2017}
Innocent~Ngaruye, Joseph~Nzabanita, D. v.~R. and Singull, M. (2017).
\newblock Small area estimation under a multivariate linear model for repeated
  measures data.
\newblock {\em Communications in Statistics - Theory and Methods},
  46(21):10835--10850.

\bibitem[Jennrich, 1969]{jennrich1969asymptotic}
Jennrich, R.~I. (1969).
\newblock Asymptotic properties of non-linear least squares estimators.
\newblock {\em Annals of Mathematical Statistics}, 40(2):633--643.

\bibitem[Jiang and Lahiri, 2001]{jiang2001empirical}
Jiang, J. and Lahiri, P. (2001).
\newblock Empirical best prediction for small area inference with binary data.
\newblock {\em Annals of the Institute of Statistical Mathematics},
  53:217--243.

\bibitem[Jiang and Lahiri, 2006]{jiang2006mixed}
Jiang, J. and Lahiri, P. (2006).
\newblock Mixed model prediction and small area estimation.
\newblock {\em Test}, 15:1--96.

\bibitem[Kim et~al., 2023]{kim2023}
Kim, J.~K., Rao, J. N.~K., and Wang, Z. (2023).
\newblock Hypotheses testing from complex survey data using bootstrap weights:
  A unified approach.
\newblock {\em Journal of the American Statistical Association},
  Accepted:1--11.

\bibitem[Kim et~al., 2018]{kim2018combining}
Kim, J.~K., Wang, Z., Zhu, Z., and Cruze, N.~B. (2018).
\newblock Combining survey and non-survey data for improved sub-area prediction
  using a multi-level model.
\newblock {\em Journal of Agricultural, Biological and Environmental
  Statistics}, 23:175--189.

\bibitem[Krewski and Rao, 1981]{krewski1981inference}
Krewski, D. and Rao, J.~N. (1981).
\newblock Inference from stratified samples: properties of the linearization,
  jackknife and balanced repeated replication methods.
\newblock {\em The Annals of Statistics}, pages 1010--1019.

\bibitem[Lahiri and Salvati, 2023]{lahiri2023nested}
Lahiri, P. and Salvati, N. (2023).
\newblock A nested error regression model with high-dimensional parameter for
  small area estimation.
\newblock {\em Journal of the Royal Statistical Society Series B: Statistical
  Methodology}, 85(2):212--239.

\bibitem[Liu et~al., 2021]{liu2021capturing}
Liu, L., Gordon, M., Miller, J.~P., Kass, M., Lin, L., Ma, S., and Liu, L.
  (2021).
\newblock Capturing heterogeneity in repeated measures data by fusion penalty.
\newblock {\em Statistics in medicine}, 40(8):1901--1916.

\bibitem[Liu et~al., 2022]{liu2022fusion}
Liu, M., Yang, J., Liu, Y., Jia, B., Chen, Y.-F., Sun, L., and Ma, S. (2022).
\newblock A fusion learning method to subgroup analysis of {A}lzheimer's
  disease.
\newblock {\em Journal of Applied Statistics}, pages 1--23.

\bibitem[Lohr and Liu, 1994]{lohr1994comparison}
Lohr, S.~L. and Liu, J. (1994).
\newblock A comparison of weighted and unweighted analyses in the national
  crime victimization survey.
\newblock {\em Journal of Quantitative Criminology}, 10(4):343--360.

\bibitem[Lumley and Scott, 2015]{lumley2015aic}
Lumley, T. and Scott, A. (2015).
\newblock Aic and bic for modeling with complex survey data.
\newblock {\em Journal of Survey Statistics and Methodology}, 3(1):1--18.

\bibitem[Ma and Huang, 2017]{ma2017concave}
Ma, S. and Huang, J. (2017).
\newblock A concave pairwise fusion approach to subgroup analysis.
\newblock {\em Journal of the American Statistical Association},
  112(517):410--423.

\bibitem[Ma et~al., 2020]{ma2020exploration}
Ma, S., Huang, J., Zhang, Z., and Liu, M. (2020).
\newblock Exploration of heterogeneous treatment effects via concave fusion.
\newblock {\em International Journal of Biostatistics}, 16(1).

\bibitem[Marhuenda et~al., 2017]{marhuenda2017poverty}
Marhuenda, Y., Molina, I., Morales, D., and Rao, J. (2017).
\newblock Poverty mapping in small areas under a twofold nested error
  regression model.
\newblock {\em Journal of the Royal Statistical Society Series A: Statistics in
  Society}, 180(4):1111--1136.

\bibitem[Molina and Rao, 2010]{molina2010small}
Molina, I. and Rao, J.~N. (2010).
\newblock Small area estimation of poverty indicators.
\newblock {\em Canadian Journal of statistics}, 38(3):369--385.

\bibitem[Newey and McFadden, 1994]{engle1994handbook}
Newey, W.~K. and McFadden, D. (1994).
\newblock Large sample estimation and hypothesis testing.
\newblock {\em Handbook of Econometrics}, 4:2111--2245.

\bibitem[Pfeffermann, 1993]{pfeffermann1993role}
Pfeffermann, D. (1993).
\newblock The role of sampling weights when modeling survey data.
\newblock {\em International Statistical Review}, 61(2):317--337.

\bibitem[Pfeffermann and Sverchkov, 2007]{Pfeffermann2007}
Pfeffermann, D. and Sverchkov, M. (2007).
\newblock Small-area estimation under informative probability sampling of areas
  and within the selected areas.
\newblock {\em Journal of the American Statistical Association},
  102(480):1427--1439.

\bibitem[Rand, 1971]{rand1971objective}
Rand, W.~M. (1971).
\newblock Objective criteria for the evaluation of clustering methods.
\newblock {\em Journal of the American Statistical Association},
  66(336):846--850.

\bibitem[Rao and Molina, 2015]{rao2015small}
Rao, J. N.~K. and Molina, I. (2015).
\newblock {\em Small area estimation}.
\newblock Wiley, Hoboken, 2nd edition.

\bibitem[Rockafellar, 1970]{rockafellar1970convex}
Rockafellar, R.~T. (1970).
\newblock {\em Convex Analysis}.
\newblock Princeton University Press.

\bibitem[Rojas-Perilla et~al., 2019]{Rojas2019}
Rojas-Perilla, N., Pannier, S., Schmid, T., and Tzavidis, N. (2019).
\newblock Data-driven transformations in small area estimation.
\newblock {\em Journal of the Royal Statistical Society Series A: Statistics in
  Society}, 183(1):121--148.

\bibitem[Rubin-Bleuer and Kratina, 2005a]{rubin2005two}
Rubin-Bleuer, S. and Kratina, I.~S. (2005a).
\newblock On the two-phase framework for joint model and design-based
  inference.
\newblock {\em Annals of Statistics}, 33(6):2789--2810.

\bibitem[Rubin-Bleuer and Kratina, 2005b]{Rubin_Kratina}
Rubin-Bleuer, S. and Kratina, I.~S. (2005b).
\newblock {On the two-phase framework for joint model and design-based
  inference}.
\newblock {\em The Annals of Statistics}, 33(6):2789 -- 2810.

\bibitem[Sun et~al., 2022]{sun2022bivariate}
Sun, H., Berg, E., and Zhu, Z. (2022).
\newblock Bivariate small-area estimation for binary and gaussian variables
  based on a conditionally specified model.
\newblock {\em Biometrics}, 78(4):1555--1565.

\bibitem[Sun et~al., 2024]{sun2024multivariate}
Sun, H., Berg, E., and Zhu, Z. (2024).
\newblock Multivariate small-area estimation for mixed-type response variables
  with item nonresponse.
\newblock {\em Journal of Survey Statistics and Methodology}, 12(2):320--342.

\bibitem[Tang and Song, 2016]{tang2016fused}
Tang, L. and Song, P.~X. (2016).
\newblock Fused lasso approach in regression coefficients clustering: learning
  parameter heterogeneity in data integration.
\newblock {\em The Journal of Machine Learning Research}, 17(1):3915--3937.

\bibitem[Wang et~al., 2007]{wang2007tuning}
Wang, H., Li, R., and Tsai, C.-L. (2007).
\newblock Tuning parameter selectors for the smoothly clipped absolute
  deviation method.
\newblock {\em Biometrika}, 94(3):553--568.

\bibitem[Wang et~al., 2014]{WANG2014409}
Wang, L., Wang, S., and Wang, G. (2014).
\newblock Variable selection and estimation for longitudinal survey data.
\newblock {\em Journal of Multivariate Analysis}, 130:409--424.

\bibitem[Wang, 2024]{wang2023clustering}
Wang, X. (2024).
\newblock Clustering of longitudinal curves via a penalized method and em
  algorithm.
\newblock {\em Computational Statistics}, 39:1485–1512.

\bibitem[Wang et~al., 2023a]{wang2023clustered}
Wang, X., Zhang, X., and Zhu, Z. (2023a).
\newblock Clustered coefficient regression models for poisson process with an
  application to seasonal warranty claim data.
\newblock {\em Technometrics}, 65(4):514--523.

\bibitem[Wang and Zhu, 2019]{wang2019small}
Wang, X. and Zhu, Z. (2019).
\newblock Small area estimation with subgroup analysis.
\newblock {\em Statistical Theory and Related Fields}, 3(2):129--135.

\bibitem[Wang et~al., 2023b]{wang2019spatial}
Wang, X., Zhu, Z., and Zhang, H.~H. (2023b).
\newblock Spatial heterogeneity automatic detection and estimation.
\newblock {\em Computational Statistics \& Data Analysis}, 180.

\bibitem[Zhang, 2010]{zhang2010nearly}
Zhang, C.-H. (2010).
\newblock Nearly unbiased variable selection under minimax concave penalty.
\newblock {\em Annals of statistics}, 38(1):894--942.

\bibitem[Zhang and Chambers, 2004]{Zhang2004}
Zhang, L.-C. and Chambers, R.~L. (2004).
\newblock {Small Area Estimates for Cross-Classifications}.
\newblock {\em Journal of the Royal Statistical Society Series B: Statistical
  Methodology}, 66(2):479--496.

\bibitem[Zhang et~al., 2022]{zhang2022subgroup}
Zhang, X., Zhang, Q., Ma, S., and Fang, K. (2022).
\newblock Subgroup analysis for high-dimensional functional regression.
\newblock {\em Journal of Multivariate Analysis}, 192:105100.

\bibitem[Zhao et~al., 2022]{zhao2022sample}
Zhao, P., Haziza, D., and Wu, C. (2022).
\newblock Sample empirical likelihood and the design-based oracle variable
  selection theory.
\newblock {\em Statistica Sinica}, 32(1):435--457.

\bibitem[Zhu and Qu, 2018]{zhu2018cluster}
Zhu, X. and Qu, A. (2018).
\newblock Cluster analysis of longitudinal profiles with subgroups.
\newblock {\em Electronic Journal of Statistics}, 12(1):171--193.

\bibitem[Zhu et~al., 2021]{zhu2021longitudinal}
Zhu, X., Tang, X., and Qu, A. (2021).
\newblock Longitudinal clustering for heterogeneous binary data.
\newblock {\em Statistica Sinica}, 31(2):603--624.

\end{thebibliography}
\end{document}